\providecommand{\sorthelp}[1]{}

\documentclass{aa}
\usepackage[utf8]{inputenc}
\usepackage{graphicx}
\usepackage{longtable}
\usepackage{booktabs}
\usepackage{float}
\usepackage{dcolumn}
\usepackage{graphics,epsfig}
\usepackage{amsmath,amssymb,latexsym,mathrsfs}
\usepackage{bm}
\usepackage[dvipsnames]{xcolor}
\usepackage{color}
\usepackage{siunitx}

\def\Planck{\textit{Planck}}

\begin{document}

   \title{PACT I: Combining ACT and {\it Planck} data for optimal detection of tSZ signal} 

   \author{N. Aghanim\inst{1} \and M. Douspis\inst{1} \and G. Hurier\inst{2} \and D. Crichton\inst{3} \and J.-M. Diego\inst{4} \and M. Hasselfield\inst{5,6} \and J. Macias-Perez\inst{7} \and T.A.~Marriage\inst{8} \and E. Pointecouteau\inst{9} \and M. Remazeilles\inst{10} \and E. Soubri\'e\inst{1} 
          }

\institute{Institut d'Astrophysique Spatiale, CNRS (UMR8617) Universit\'{e} Paris Sud, B\^{a}timent 121, Orsay, France
\and Centro de Estudios de F\'isica del Cosmos de Arag\'on (CEFCA),Plaza de San Juan, 1, planta 2, E-44001, Teruel, Spain 
\and Astrophysics and Cosmology Research Unit, School of Mathematics, Statistics and Computer Science, University of KwaZulu-Natal, Durban 4041, South Africa
\and Instituto de F\'isica de Cantabria (CSIC-UC). Avda. Los Castros s/n. 39005 Santander, Spain
\and Department of Astronomy and Astrophysics, The Pennsylvania State University,
   University Park, PA 16802, USA
\and Institute for Gravitation and the Cosmos, The Pennsylvania State University,
   University Park, PA 16802, USA
\and LPSC, Université Joseph Fourier Grenoble 1, CNRS/IN2P3, Institut National Polytechnique de Grenoble, 53, av. des Martyrs, 38026 Grenoble, France
\and Dept. of Physics and Astronomy, Johns Hopkins University, 3400 N. Charles St., Baltimore, MD 21218, USA
\and IRAP, Université de Toulouse, CNRS, CNES, UPS, (Toulouse), France
\and Jodrell Bank Centre for Astrophysics, School of Physics and Astronomy, The University of Manchester, Manchester M13 9PL, UK\\
\email{nabila.aghanim@ias.u-psud.fr} 
}

   \date{Received /Accepted}
 
   \abstract{We present the optimal reconstruction of the  thermal Sunyaev-Zel'dovich (tSZ) effect signal based on the combination of a heterogeneous dataset consisting of ACT and  {\it Planck} data, with different numbers of channels, angular resolutions and noise levels. We combine both datasets using two different approaches, a Matched Multi-Filter (MMF) technique and an optimised Internal  Linear  Combination (ILC). We show that when applying the MMF to the combination of ACT and {\it Planck} data, the size-flux degeneracy is reduced and the signal-to-noise of clusters detected with their SZ signal improves by up to a factor of three. In the case of the optimised ILC method, we show that the tSZ map is reconstructed with a resolution of $\sim 1.5$ arcmin. This is more than a factor two improvement compared with the {\it Planck} resolution, and  with a very good control of noise, i.e. limited only by the intrinsic noise of the individual experiments. The combination of ACT and  {\it Planck} data offers a unique opportunity to improve on the study of the pressure profiles and to study substructure in clusters through their tSZ.}    
  \maketitle



\section{Introduction}

The thermal Sunyaev-Zel'dovich (tSZ) effect is a spectral distortion of the  Cosmic Microwave Background (CMB) black body radiation due to inverse Compton scattering of the CMB photons by hot electrons contained in the intra-cluster gas \citep{Zeldovich1969,1999PhR...310...97B}. In the non-relativistic approximation, the tSZ effect has a distinctive frequency signature independent of redshift. Furthermore, for clusters of galaxies (dominant contribution to the tSZ signal) there  is also a distinctive spatial distribution associated with their pressure profile, which translates into a dependency also in the angular power spectrum of the tSZ. The microwave sky is the result of several astrophysical, and cosmological signals including the CMB, foreground  emission  from  the  Galactic  interstellar medium (ISM), the Cosmic Infrared Background (CIB) emission, the emission from compact extra-galactic radio sources and the aforementioned tSZ. These signals have usually distinctive frequency signatures and spatial distribution on the sky. In addition to these foregrounds, observations also include  instrumental noise. Separating the different astrophysical and cosmological components is a challenging task, but in general, this separation improves when more frequencies are added into the reconstruction method. 

Thanks to the recent CMB surveys by the Atacama Cosmology Telescope (ACT) \citep{2011ApJ...737...61M,2013JCAP...07..008H,2018ApJS..235...20H}, South Pole Telescope (SPT) \citep{2009ApJ...701...32S,2015ApJS..216...27B}, and {\it Planck} \citep[][ and references therein]{2018arXiv180706205P}, wide-field multi-frequency data are now available for tSZ studies. These surveys differ notably in terms of their numbers of frequency channels and in terms of their angular resolutions from $\sim 1.5$ arcmin for ground-based surveys to $\sim 5$ arcmin for the highest {\it Planck} frequencies.  Also, the noise properties can vary significantly, especially for the ground-based experiments, where changes in the atmospheric conditions can result in spatial variations at large angular scales.
In this context,  reconstruction of the tSZ signal relies on our capacity to optimally combine the available multi-frequency and multi-instrument data. 

Several component separation methods have been successfully used in the {\it Planck} context \citep[e.g.,][]{2008StMet...5..307B,2011MNRAS.418..467R,2013A&A...558A.118H}.
In \cite{2013MNRAS.430..370R} a first study based on simulated data showed that joint exploitation of multiple experiments producing heterogeneous data is possible. 
The needlet Internal  Linear  Combination (ILC) method used in \cite{2013MNRAS.430..370R} efficiently aggregates simulated ACT-like and {\it Planck}-like data to construct high resolution tSZ patch-maps around clusters of galaxies.  
The first combination of data from multiple heterogeneous CMB experiments to produce an optimal temperature map was performed by \cite {2014A&A...563A.105B,2016A&A...591A..50B} using the {\it Planck} and WMAP data. 
A more recent study by \cite{2018ApJS..239...10C} produced a CMB map from the linear combination of SPT and {\it Planck} data. {\it Planck} and SPT data were also combined to produce maps of the Magellanic
Clouds \citep{2016ApJS..227...23C}. 

In this paper, we present the first optimal reconstruction of the tSZ effect signal based on the combination of  publicly available ACT and {\it Planck} data. The combination based on an ILC approach takes advantage of the complementarity between the two experiments: the higher resolution of ACT to detect compact tSZ clusters and characterize pressure profiles on smaller angular scales and the larger number of frequencies and measurement stability of {\it Planck} to clean the foreground contamination and CMB and allow access to larger angular scales filtered out in ground experiments. The paper is organised as follows. In Sect. \ref{sec:data}, we provide a brief description of the data used. We then present two ways of combining ACT and {\it Planck} data -- the Matched Multi-Filter used to optimally detect clusters of galaxies and the ILC method to reconstruct a tSZ map -- in Sect. \ref{sec: combi}. The results in terms of the tSZ radial profile, significance of tSZ detection and size-flux measurement for clusters are reported in Sect. \ref{sec:prop}. We conclude in Sect. \ref{sec:ccl}.


\section{Data}\label{sec:data}

We focus on two CMB surveys, ACT and {\it Planck}, which differ by their angular resolutions, by their available frequency channels, and by their sky coverage. In the following, both datasets are considered as a single composite dataset of intensity and associated noise maps. Characteristics of the maps from both experiments are summarized in Table~\ref{tab:char}.
We perform our analysis using this composite multi-instrument data that will be referred to as the PACT dataset in the rest of the paper.

We use the publicly available total intensity dataset of the High Frequency Instrument (HFI) taken during the full mission of the \Planck\ survey, i.e. the six highest-frequency {\it
 Planck} channel maps, from 100 to 857~GHz at native resolution \citep{planck2014-a09}. The full {\it Planck} dataset is available at the {\it Planck} Legacy Archive (PLA\footnote{http://www.sciops.esa.int}). A noise map is
associated with each channel map and is constructed from the
difference of the first half and second half of the {\it Planck} rings for a given pointing position of the satellite spin axis. These maps
are mainly free from astrophysical emission and they are
considered a good representation of the statistical instrumental noise
\citep{planck2014-a09}. As the goal of this work is to focus on SZ signal, in addition to the intensity maps we also use publicly available mask maps constructed to remove regions of the sky affected by point sources and strong galactic emission, also available at PLA. 

We use {\it Planck} circular Gaussian beams with FWHM values
from \citep[][Table 3]{planck2014-a08} and the values provided
in \cite{planck2014-a28} for the tSZ transmission in
{\it Planck} spectral bandpasses. To allow a direct combination of {\it
Planck} and ACT datasets, all {\it Planck} maps were projected into
the world coordinate system of the 148~GHz ACT map.

For the ACT dataset, we use the publicly available 148 and 220~GHz data in the Southern and Equatorial areas at native
resolution 1--1.4 arcmin. The data was collected from 2008--2010 using the Millimeter Bolometric Array Camera \citep{2011ApJS..194...41S}. The ACT data reduction is described in \cite{2013ApJ...762...10D}. We also use the associated weight maps and point-source masks\footnote{All ACT
data products can be found at the Legacy Archive for Microwave Background Data (https://lambda.gsfc.nasa.gov/).}. The latter are defined to be zero in a 5 arcmin circle at the location of sources detected with flux density greater than 15 mJy at 148~GHz.\footnote{Note that no source detection was performed in the 18-hour field of the Southern area.}  Finally, we use the beam characteristics and tSZ transmission band centers in \cite{2013ApJS..209...17H} and \cite{2011ApJS..194...41S}, respectively.

In the following, we combine the full ACT dataset with the overlapping {\it Planck} coverage. However, we identify two deeper regions that will be used to illustrate the results, called S$_{\mathrm{deep}}$ and E$_{\mathrm{deep}}$ for the Southern and Equatorial regions, respectively. These selected areas are displayed as black rectangles in Fig.~\ref{fig:noisemaps}. The area S$_{\mathrm{deep}}$ is defined such that ${56.3^\circ<\mathrm{R.A.}<97.9^\circ}$  and ${-55.3^\circ<\mathrm{ Dec.}<-51.2^\circ}$. The area  E$_{\mathrm{deep}}$ is such that ${4.1^\circ<\mathrm{R.A.}<53.6^\circ}$  and ${-1.06^\circ<\mathrm{Dec.}<1.40^\circ}$.

\begin{table}
\caption{\label{tab:char}{\it Planck} and ACT frequencies, beams and SZ conversion factors. The conversion factors are computed without relativistic corrections.}
\centering
\begin{tabular}{lccr}
\hline\hline
\noalign{\vskip 2pt}
    &Frequency&FWHM  & g($\nu$) $T_{\mathrm{CMB}}$  \\
    & [GHz] & [arcmin] & [$K_{\mathrm{CMB}}$]\\
\noalign{\vskip 2pt}
\hline
\noalign{\vskip 2pt}
{\it Planck}    & 100     &9.68&-4.03121 \\
          & 143     &7.30&-2.78564\\
          & 217     &5.02& 0.18763\\
          & 353     &4.94& 6.20518\\
          & 545     &4.83&14.45559\\
          & 857     &4.64&26.33521\\
\noalign{\vskip 2pt}     
\hline
\noalign{\vskip 2pt}
ACT       &148     &1.374 &-2.69100\\
          &220     &1.053 &0.10400\\
\noalign{\vskip 2pt}
\hline
\end{tabular}
\end{table}


\section{Combining ACT and {\it Planck} data}\label{sec: combi}

In this section, we present two different approaches for the combination of the ACT and {\it Planck} datasets, both optimised for the tSZ studies. The first one consists in combining the ACT and {\it Planck}  maps through a Matched Multi-Filter (MMF) technique. This method is particularly optimised for the detection of tSZ sources and the measurement of their significance and photometry. The second approach consists in reconstructing a tSZ $y$-map using an ILC method to combine the frequency maps of the PACT dataset.

We show, in Sect. \ref{sec:prop}, some of the advantages of the PACT dataset over the individual {\it Planck} and ACT data. Full analyses of the tSZ source catalog and $y$-map derived from the PACT dataset are beyond the scope of the present study. Hence as a proof of concept, we focus mostly on the deepest region of the ACT data, S$_{\mathrm{deep}}$, and we illustrate our results and compare them with those obtained, for the same region, by {\it Planck} and ACT separately.

\begin{figure*}[!th]
\begin{center}

\includegraphics[scale=0.18]{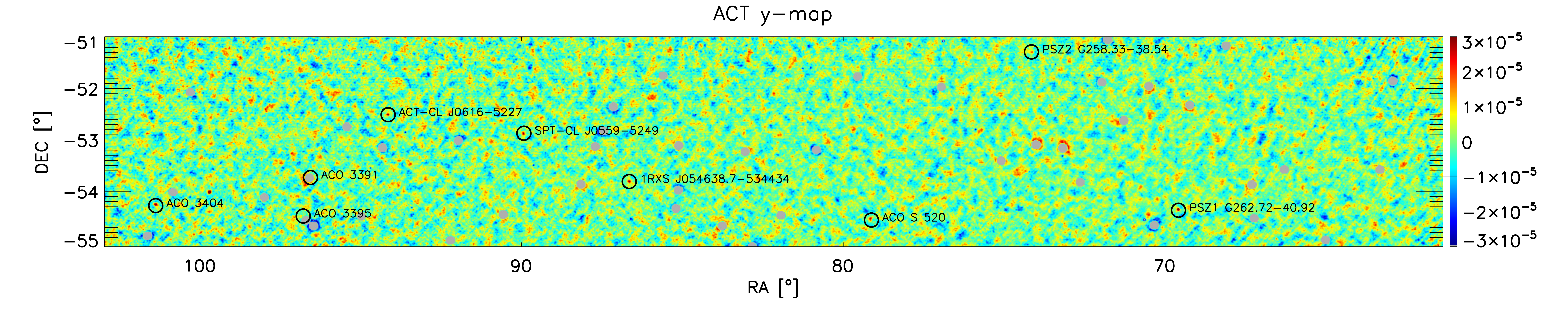}
\includegraphics[scale=0.18]{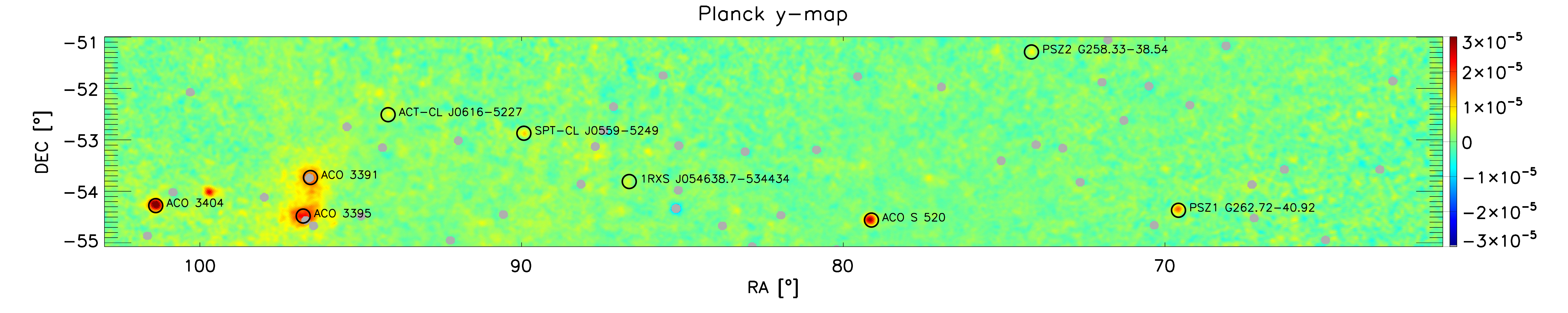}
\includegraphics[scale=0.18]{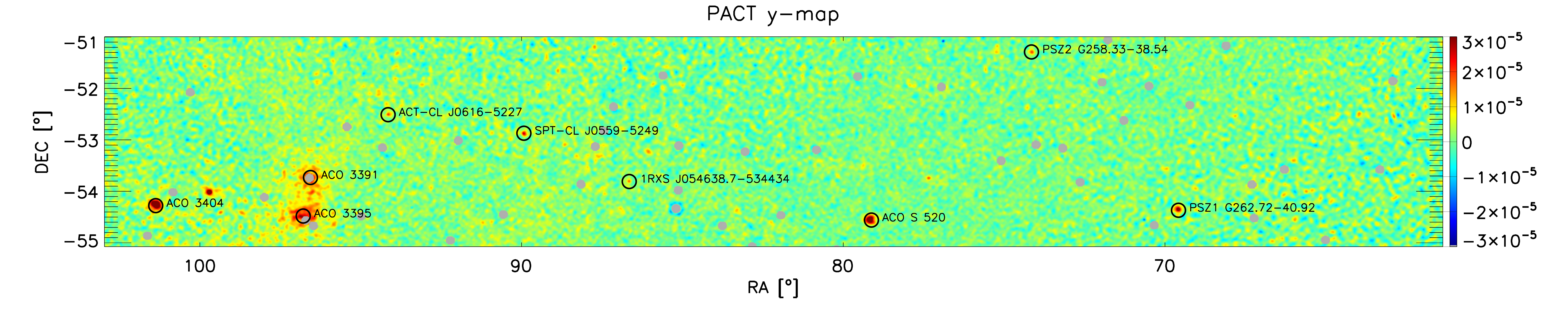}

\caption{Maps of the unit-less (color bar) Compton $y$ parameter illustrated on the selected Southern ACT area. Top panel: 148GHz-ACT map at 1.5 arcmin resolution with scales larger than 7 arcmin filtered out. Middle panel: {\it Planck} $y$-map reconstructed with MILCA at 7 arcmin resolution. Lower panel: PACT $y$-map reconstructed with MILCA at 3 arcmin resolution. Circles indicate some known clusters in the selected area. The ACT point sources are masked in all three panels (grey areas). }
\label{PACT_south}
\end{center}
\end{figure*}

\subsection{Matched Multi-Filter for PACT data}\label{sec:cat}
Several detection methods geared towards the detection of tSZ clusters were proposed in the literature. Matched Multi-Filter (MMF) techniques \citep[e.g.,][and references therein]{2002MNRAS.334..533H,melin2006,carvalho2009} have been successfully used in the context of both single- and multi-frequency CMB surveys such as ACT, SPT, and {\it Planck}. They have produced large and complementary catalogs of galaxy clusters\footnote{A meta-catalog collecting all the catalog entries of these individual experiments is 
available at http://szcluster-db.ias.u-psud.fr/.}. 

The power of the MMF technique
resides in the fact that it enhances the signal-to-noise (S/N) of a tSZ source by optimally filtering a set of multi-frequency sky maps containing an ensemble of contaminating signals (e.g., galactic dust,
CMB, etc.). The optimal filtering of the tSZ signal is achieved thanks to two characteristics. On the one hand, the tSZ spectral signature is
assumed to be known and universal, as far as the electrons of the intracluster medium are non-relativistic. On the other hand, the tSZ cluster signal is described by a spatial template following a projected spherical
Generalised Navarro-Frenk-White profile \citep{arn10}
characterised by a single parameter: a characteristic radius
$\theta_{\mathrm{s}}$. The optimal filter defined this way rejects both the foreground contamination and the instrumental noise, using a linear combination of the frequency maps.

Being a detection method based on matched filters for multiple frequency data, the MMF is particularly well adapted for the combination of the ACT and {\it Planck} data. In the present study, we use all six frequency maps from the {\it Planck}-HFI instrument and the ACT frequency maps together with the associated noise maps at their native resolutions.

The MMF is specifically designed to reject foreground contamination however it was shown
in \citet{planck2011-5.1a,planck2013-p05a,planck2013-p05a-addendum} and \cite{planck2014-a36} that spurious SZ sources can be associated with IR emission from cold galactic clumps or other point-like sources. Many strategies can be followed in order to clean the tSZ detections from spurious sources such as cross-matching with external cluster catalogs, comparison with galaxy surveys or statistically-based classification \citep{2015A&A...580A.138A}. One can at the same time attempt to reduce the possible contaminations at an earlier stage of the analysis. To do so, we apply to the ensemble of PACT frequency maps, before their analysis, a mask for point sources consisting of the union of the \Planck\ and ACT point-sources masks presented in Sect. \ref{sec:data}. The PACT dataset, i.e. all {\it Planck}-HFI and ACT channel maps, is then processed with an MMF filter. For this, 
the size of the spatial template is varied between 0.1 and 50 arcmin in 50 logarithmically spaced bins. A detected tSZ source is defined as the one with the highest
S/N when varying the template size. The associated size and flux measured from the filter are defined as the ones of detected source. 

It is beyond the scope of the present study to construct and deliver a catalog of clusters and tSZ sources from the PACT dataset. Our aim is to show the complementarity and added value of combining ACT and {\it Planck} data. We thus illustrate the
application of the MMF to the detection of the tSZ sources in the
combined PACT dataset by focusing on results of the deep region S$_{\mathrm{deep}}$. 
Furthermore, we have checked neither S/N
nor tSZ fluxes derived from the MMF method are improved when we use both the 148 and the 220~GHz ACT
maps or only the 148~GHz map. We therefore choose to show the results of the analysis using all eight frequency channels of the PACT dataset (the two from ACT and six from Planck). 

In this configuration, we detect in total 58 tSZ
sources above a threshold of S/N=4.3. This threshold corresponds to the lowest signal-to-noise of the tSZ sources identified with actual clusters (i.e. below S/N=4.3 the tSZ sources are either candidate tSZ clusters or spurious detections). The sample of 58 tSZ PACT sources contains 22 actual clusters. The identification of counterparts is performed by searching for already-known clusters from tSZ, X-ray and optical catalogs in a 5 arcmin radius around the position of the PACT detected tSZ sources. We use the tSZ compilation of 2690 tSZ sources\footnote{It is based on the catalogs from {\it Planck}, ACT, SPT, CARMA, AMI. The meta-catalog is
available at http://szcluster-db.ias.u-psud.fr/}. We also use the MCXC meta-catalog \citep{2011A&A...534A.109P}. In the optical, we use catalogs from the SDSS survey \citep{Rykoff2014,Wen2015}. Finally, we query the NED database for the PACT sources that are not associated with known clusters.
 \\

Thanks to the optimal combination of the ACT and {\it Planck} data with the MMF, we are able to recover all the clusters detected by ACT and {\it Planck} not only those common to both experiments but also those detected independently by ACT or by {\it Planck}. This amounts to 14 clusters of galaxies in total. Furthermore, all {\it Planck} clusters recovered
from the PACT data have $S/N>4.5$, the {\it Planck} detection threshold. We discuss later the improvement in S/N due to the PACT combination of data. The improvement brought by the PACT dataset can also be seen in the detection of eight new tSZ sources in
the S$_{\mathrm{deep}}$ region associated with actual clusters from
optical, X-ray or tSZ catalogs. Their redshifts range from $z=0.21$
to 0.93. Thanks to the PACT  dataset combining ACT's high resolution and {\it Planck}'s cleaning capabilities, we detect
clusters at $z\simeq 0.9$ that were  reported neither in the ACT nor in the {\it Planck} catalogs. This illustrates the potential of the PACT data for high redshift clusters.


\subsection{PACT $y$-map construction} \label{sec:map}

We now combine the {\it Planck} and ACT data  to build a joint tSZ $y$-map, called the PACT $y$-map. As described in \cite{2013MNRAS.430..370R}, such a combination can be performed using component separation techniques which aggregate the data. We choose here to illustrate the results for the MILCA method (see \cite{2013A&A...558A.118H} for details).

The MILCA method allows us to perform, for the whole region Equatorial and South regions covered
both by ACT and {\it Planck}, a scale dependent optimization of the reconstructed tSZ signal with respect to noise and foreground residuals.  Figure \ref{PACT_south} shows the resulting reconstructed $y$-map in a selected area of the PACT footprint. 
MILCA is a semi-blind component separation method that requires as input the tSZ spectral dependence. Additional constraints can be considered to optimize the separation. In the present case, we use a second spectral constraint to suppress the CMB contribution in the reconstructed $y$-map at large scale. At small scale (below 5 arcmin), the relevant signal is dominated by the ACT-148 GHz map, thus CMB cannot be spectrally separated from the tSZ emission. However, at such small angular scales CMB primary anisotropies can be safely neglected. Similarly, contamination by infrared and radio point sources cannot be suppressed at very small scales; they hence produce respectively spurious positive and negative sources in the MILCA-produced PACT $y$-map.

\begin{figure}[!th]
\begin{center}
\resizebox{\hsize}{!}{\includegraphics{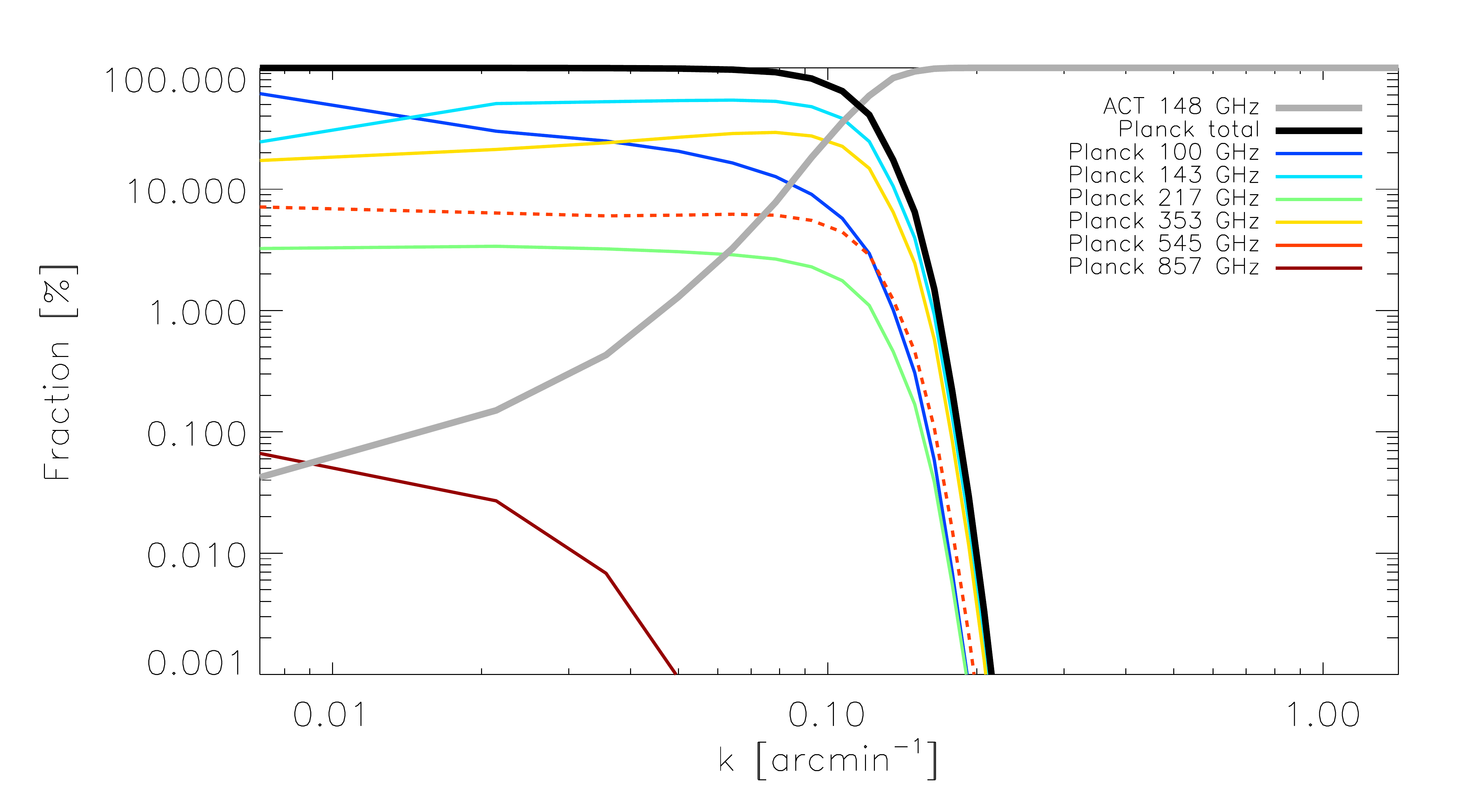}}
\caption{Relative contribution, i.e. weights, as a function of scales to the reconstructed PACT $y$-map of each
  frequency channel: 100~GHz (dark blue), 143~GHz (light blue), 148~GHz (thick grey), 217~GHz
  (green), 353~GHz (yellow), 545~GHz (dashed red line), and 857~GHz (brown). The relative contribution to the reconstructed PACT $y$-map from {\it Planck} is shown in thick black line.  }
\label{weights}
\end{center}
\end{figure}

In Fig.~\ref{weights}, we show the relative contribution, i.e. weight, from each of the seven channel/intensity maps to the PACT $y$-map. We do not include the 220~GHz channel from ACT in the reconstruction of the tSZ map by MILCA since it improves only very slightly the statistical noise (by about 1\%) whereas it introduces
correlated noise from sky background which is more difficult to model and thus to control. The weights show that ACT and {\it Planck} experiments nicely complement each other and that their combination allows us to take advantage of {\it Planck} data at scales below $\sim\,0.1$ arcmin$^{-1}$ and ACT data above, as will be discussed below. The 148~GHz map from ACT (thick solid grey line) dominates the contribution to the PACT $y$-map at scales above $\sim\,0.1$ arcmin$^{-1}$ due to the higher ACT resolution. 

We note that most of the contribution to the tSZ signal at large angular scales in the PACT $y$-map reconstruction actually comes from the {\it Planck} frequency channels with large beams. Since they dominate the cleaning from foregrounds in the MILCA method (and more generally in Internal Linear Combination approaches). Figure~\ref{weights} highlights the dominant contributions to the PACT $y$-map from the 143 and 353~GHz channels at large scales. It also shows that the highest frequency, 857~GHz, tracing the Galactic dust emission does not contribute significantly to the reconstruction of the tSZ map. The two highest frequencies, 545 and 857~GHz, are however essential for thermal dust emission removal. Considering the sky is completely dominated by the thermal dust emission at these frequencies, and in particular at 857~GHz, only a small weight is needed for the reconstruction of the tSZ signal.

We produce a reconstructed PACT $y$-map from the combination of seven frequency maps, corresponding to the six {\it Planck} highest-frequency channels and the 148~GHz channel from ACT. We show, in Fig. \ref{PACT_south} bottom panel, the resulting $y$-map in the S$_{\mathrm{deep}}$ area. For the same region, the {\it Planck} $y$-map and the 148~GHz-ACT map, in units of Compton parameter, are displayed in the top and middle panels respectively. We clearly notice the lack of signal at large scales in the ACT $y$-map. This is due to the high-pass filtering at 7 arcmin applied to enhance tSZ-scale signal. 
The ACT $y$-map also exhibits some radio-source contamination mainly shown as a negative signal. We note that the counterpart of the brightest radio-source in the ACT $y$-maps displays a positive ring introduced by the MILCA multi-scale reconstruction that uses the various frequencies with different relative weights due to their different angular resolution (see Fig. \ref{weights}). The {\it Planck} $y$-map shows large scale tSZ scale signal due to the larger {\it Planck} beams. It also shows contamination from radio-sources at a lower amplitude than ACT $y$-map, due both to the large frequency coverage which eases the component separation and to the beam smearing. If we now turn to the PACT $y$-map (Fig.~\ref{PACT_south} bottom panel), we see the advantages of combining the two experiments into a single dataset with MILCA. 
As seen in the area of the the pair of clusters A3395-A3391, we do recover in the PACT $y$-map the large angular scales thanks to {\it Planck}. We also clearly see the improvement brought by the higher resolution of ACT both in terms of spatial structure and in terms of significance. This is seen on the two {\it Planck} clusters in the field PSZ1~G258.33-38.54 and PSZ2~G262.72-40.92. Finally, the advantage of the combination of ACT and {\it Planck} is also illustrated by the detection of a higher number of
tSZ sources seen as positive sources in the $y$-map (including actual clusters not detected independently by ACT or by {\it Planck}). The PACT reconstructed $y$-map also shows contamination by radio-sources which can be easily mitigated thanks to their negative signal. 

\begin{figure*}[!th]
\begin{center}
\resizebox{\hsize}{!}{\includegraphics{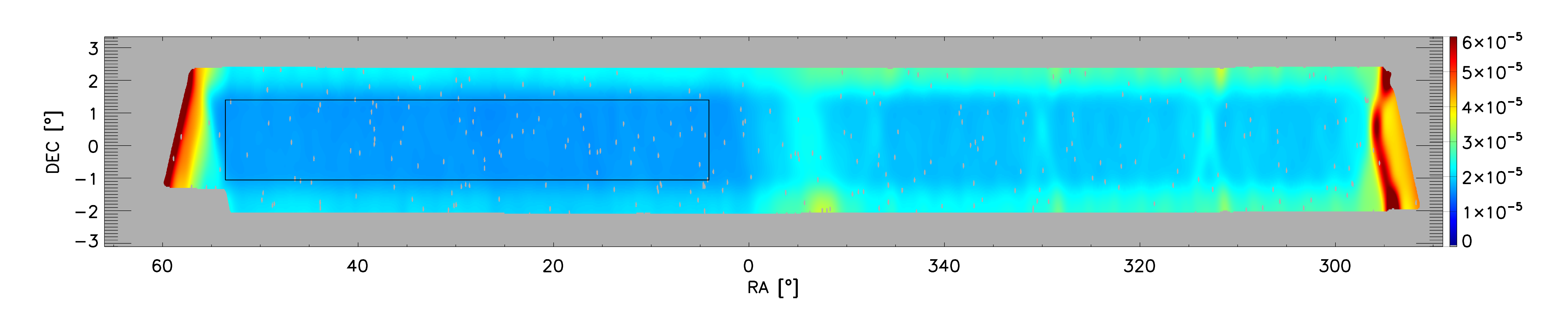}}
\resizebox{\hsize}{!}{\includegraphics{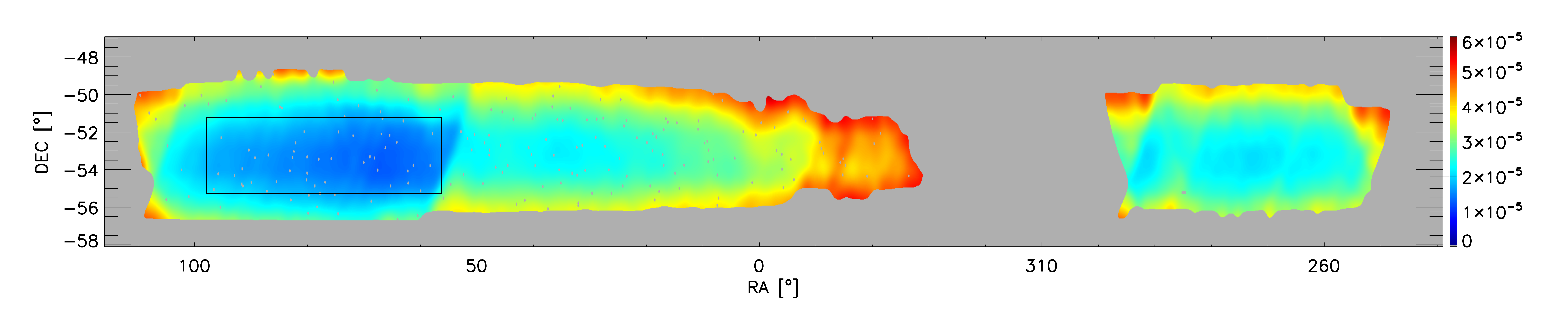}}
\caption{\label{fig:noisemaps} Combined noise maps in the native ACT resolution of 1.4 arcmin, overlaid with the ACT point-source masks. The color bar represents the unit-less Compton $y$ parameter.
Top: ACT Equatorial area. Bottom: ACT Southern area. The rectangle in each panel represents the deepest area S$_{\mathrm{deep}}$ (in the Southern region) and E$_{\mathrm{deep}}$ (in the Equatorial region).
}
\label{footprints}
\end{center}
\end{figure*}

The noise map of the PACT shown in Fig.~\ref{fig:noisemaps} is obtained from the half-dataset difference maps and applying the same linear combination (and weights) as the one used to compute the reconstructed PACT $y$-map. The PACT noise maps in Fig.~\ref{fig:noisemaps} are shown for the ACT equatorial footprint (top panel) and southern footpring (bottom panel). The rectangular regions represent selected deep regions defined in Sect. \ref{sec:data}. By construction of the PACT dataset from the combination of ACT and {\it Planck} frequency channels, the associated noise structures are scale dependent. This is exhibited in Fig. \ref{PACT_south}, bottom panel, where we note that the noise spatial distribution is correlated and inhomogeneous. Considering that large angular scales are dominated by {\it Planck} data and small angular scales by ACT data.  

\begin{figure}[!th]
\begin{center}
\resizebox{\hsize}{!}{
\includegraphics{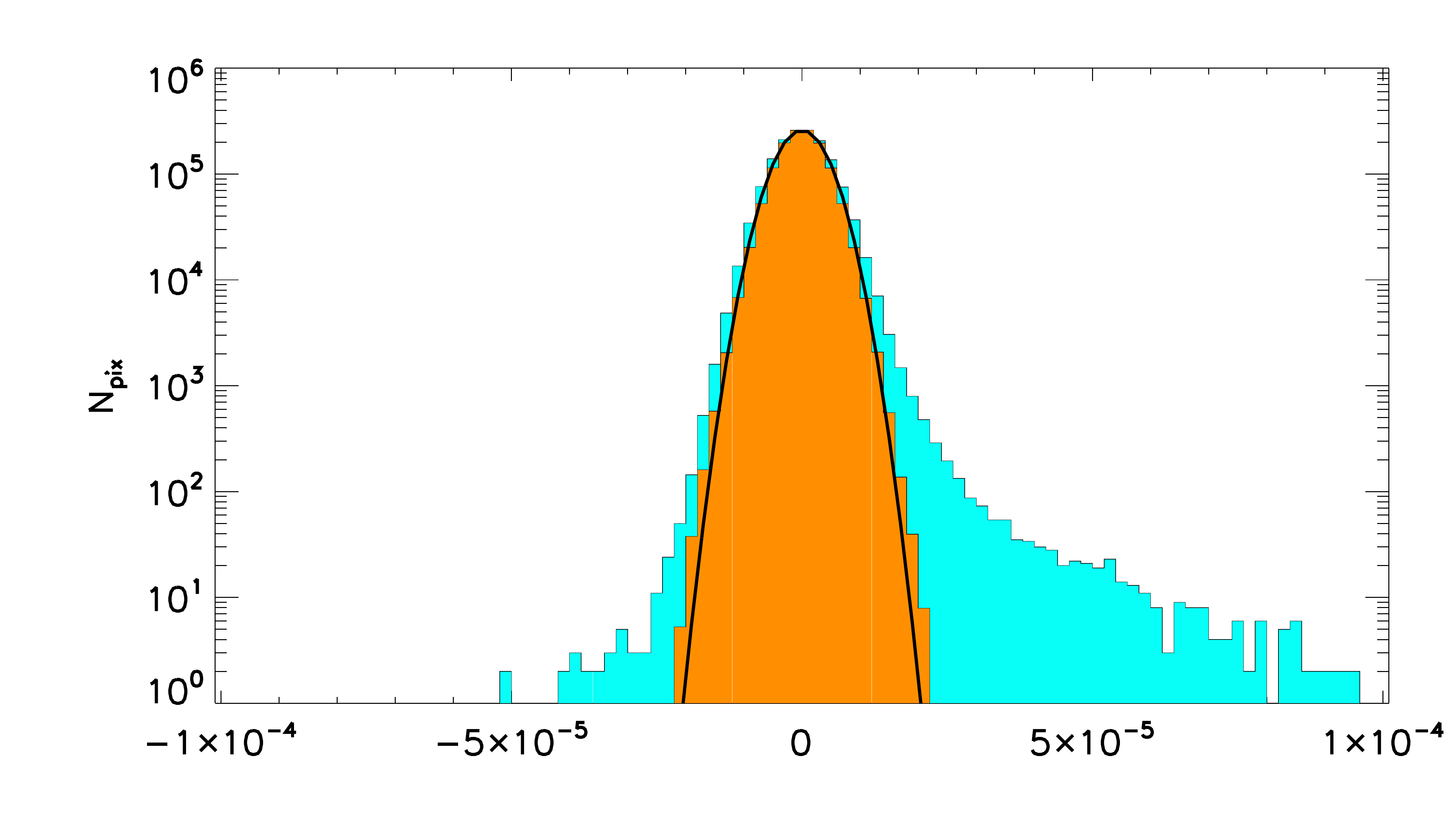}
}
\caption{Distribution of pixels for the MILCA PACT $y$-map, in cyan for the tSZ signal, and in orange for the noise, the solid black line shows a Gaussian fit adjusted to the noise distribution.}
\label{PACT_1pdf}
\end{center}
\end{figure}

To illustrate the quality of the PACT MILCA tSZ $y$-map, we present on Fig.~\ref{PACT_1pdf} the pixel distribution of both PACT maps: tSZ signal and noise. The noise is well approximated by a Gaussian distribution whereas the distribution of tSZ signal exhibits a non-Gaussian distribution with an important positive tail and a smaller negative tail.
Only considering the negative pixels in the PACT $y$-map, the background standard deviation is $4.6 \times 10^{-6}$ in Compton parameter units. The statistical noise contributes for $4.0\times 10^{-6}$ to the  standard deviation background of the PACT $y$-map. Thus other sources of background fluctuations account for $\simeq 2.3 \times 10^{-6}$. This additional contribution is likely produced by residual astrophysical emissions such as individual sources and Cosmic Infrared Background (CIB). The CIB residuals in the tSZ maps at large scale is mainly Gaussian \citep[see,][]{2016A&A...594A..23P}. However, at small scales the PACT map is dominated by the contribution from the 148~GHz channel in which the CIB conserves all its complexity and non-Gaussian characteristics.

\begin{figure*}[!th]
\begin{center}

\includegraphics[width=0.45\linewidth]{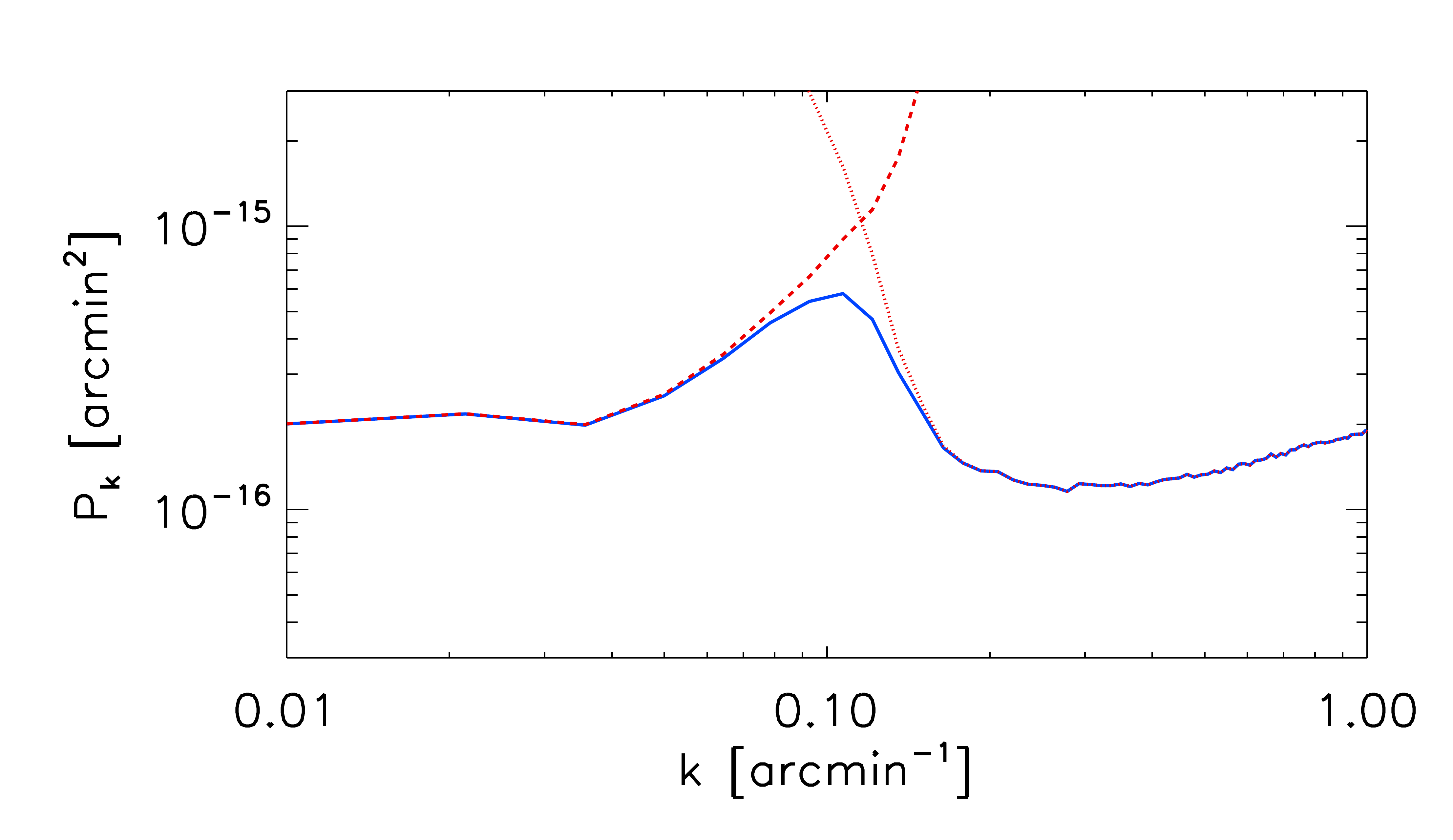}
\includegraphics[width=0.45\linewidth]{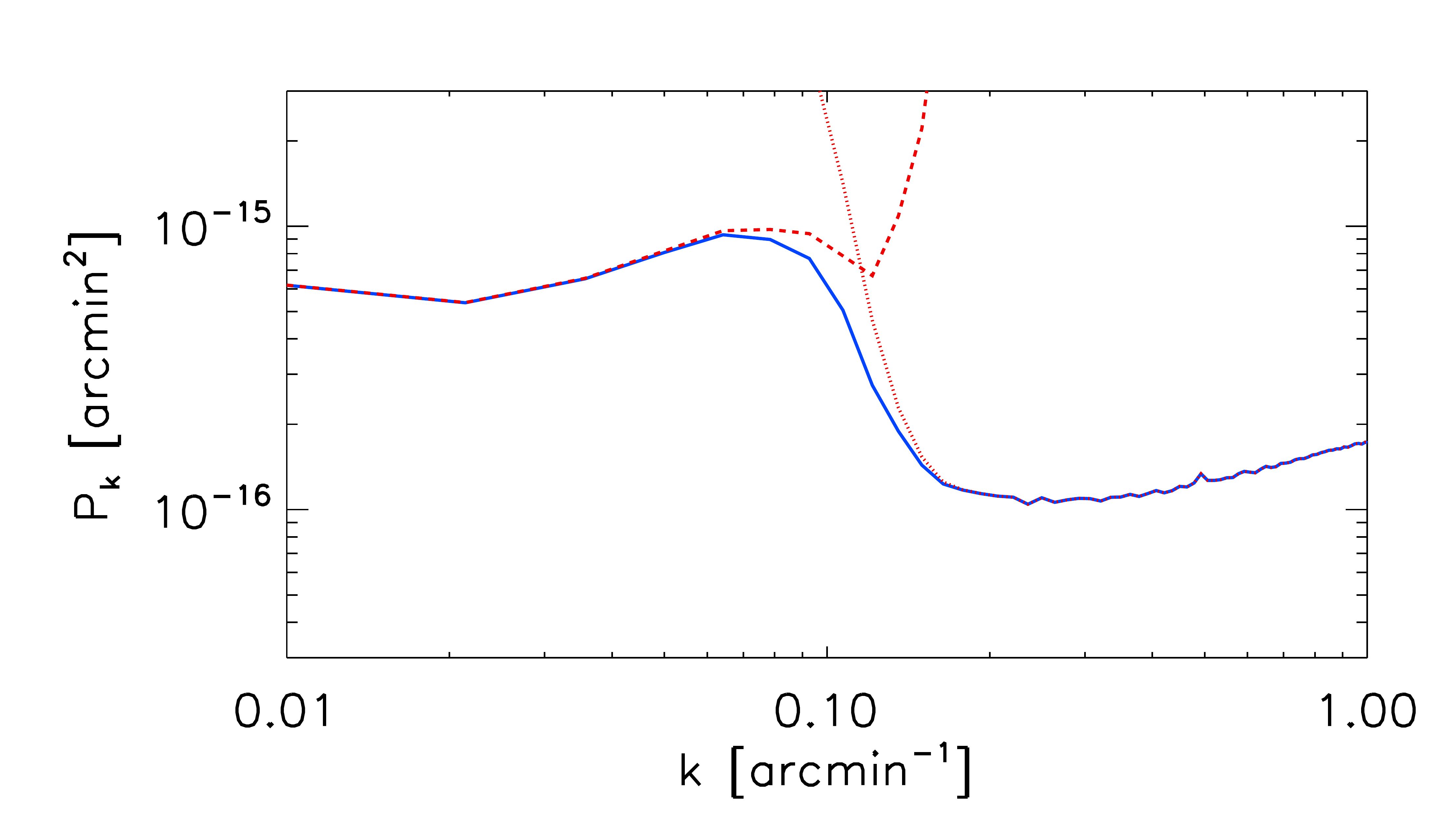}
\includegraphics[width=0.45\linewidth]{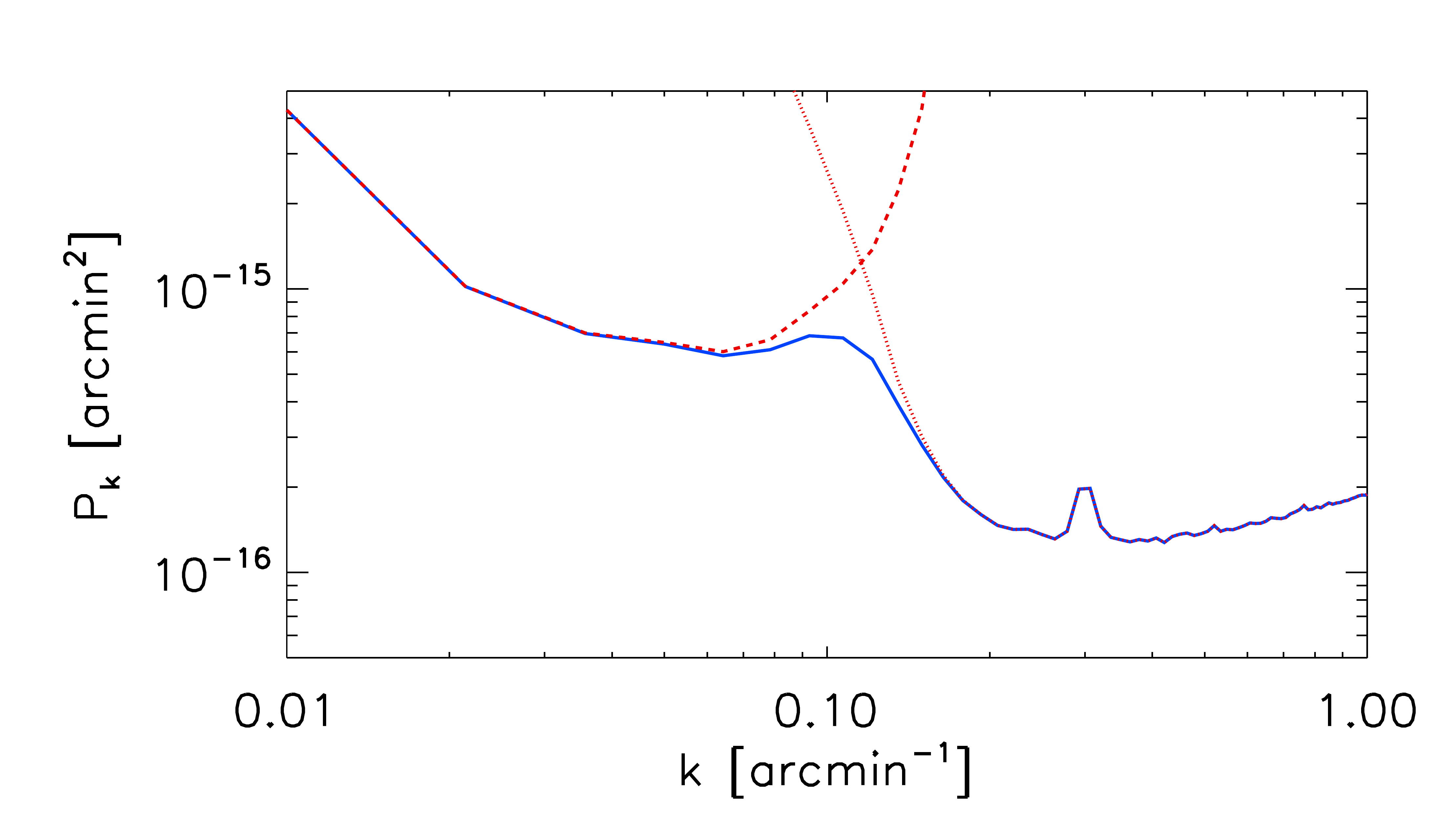}
\includegraphics[width=0.45\linewidth]{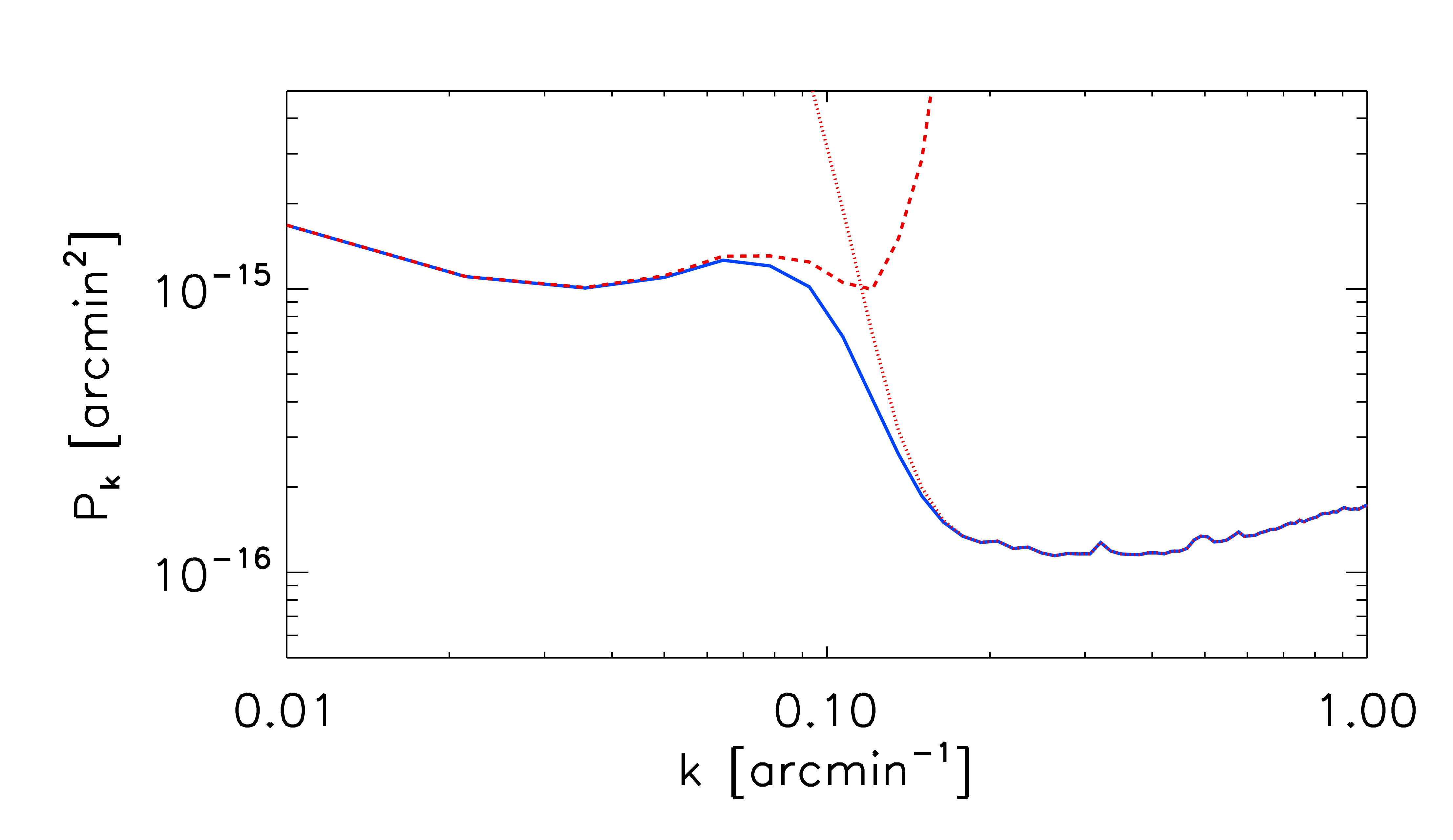}
\caption{Top: PACT tSZ noise power spectrum (solid blue) compared with {\it Planck} and ACT noise power spectrum (in red dashed and dotted line) for the Southern region (left panel) and the Equatorial region (right panel). Middle: same as top but for the tSZ signal. Bottom: R.A.-axis projected PACT (black), {\it Planck} (blue), and ACT 148~GHz (red) signals  for the same regions. }
\label{PACT_spec}
\end{center}
\end{figure*}

Finally, we compute the power spectra of the PACT $y$-map (red line in Fig. \ref{PACT_spec} middle panels) and of the associated noise map (blue line in Fig. \ref{PACT_spec} top panels) in the Southern and the Equatorial regions. 
The small-scale noise spectrum, above $k\sim0.2$arcmin$^{-1}$, is as expected dominated by that of ACT. The feature at $\sim 0.3$ arcmin$^{-1}$ in the southern PACT signal power-spectrum is associated with the stripes in the ACT 148~GHz signal map translated into oscillation in the power-spectrum that are propagated to the PACT power-spectrum.
At larger scales the noise spectrum is dominated by {\it Planck} with an amplitude slightly larger than the ACT noise in the Southern area and significantly larger in the Equatorial region due {\it Planck} scanning strategy with lower redundancy (hence higher noise) in the Equatorial region.  
We observe a large bump around 0.1 arcmin$^{-1}$ which reflects the angular scales where the contribution to the tSZ signal transits from ACT to {\it Planck}. In the Southern area, we note that the tSZ signal exhibits a clear excess of power compared to the noise power spectrum. This signal includes contributions from tSZ, CIB, and point sources \citep[see,][for a detailed analysis of the tSZ power spectrum]{2014A&A...571A..21P}.


\section{Properties of the tSZ clusters from the PACT dataset}\label{sec:prop}

The combination of the ACT and {\it Planck} data into the PACT dataset improves the yield of detected tSZ sources as discussed in Sect. \ref{sec:cat}. It also improves the quality of the reconstructed $y$-map as seen in Sect. \ref{sec:map}. We now show the improvements brought by the ACT and {\it Planck} combination focusing on the significance of the tSZ clusters detected by ACT or {\it Planck}, namely their signal to noise, on their estimated flux and sizes, and finally on the Compton parameter profile.

\subsection{Sample description}

We focus  on a union sample of 119 galaxy clusters  detected either by {\it Planck} or by
ACT. We use the ACT catalog from \cite{2013JCAP...07..008H} and the Planck catalogs from \citet{planck2011-5.1a,planck2013-p05a,planck2013-p05a-addendum} and \cite{planck2014-a36}. We identify the clusters detected either only in the ACT data or only in the {\it Planck} survey. For these clusters positions and native S/N values are those published by both experiments. We also identify the clusters detected both in ACT and in {\it Planck}. For these clusters, we choose the ACT position as a reference tSZ position given the higher resolution of this experiment. 

\begin{figure}[!th]
\begin{center}
\resizebox{\hsize}{!}{\includegraphics{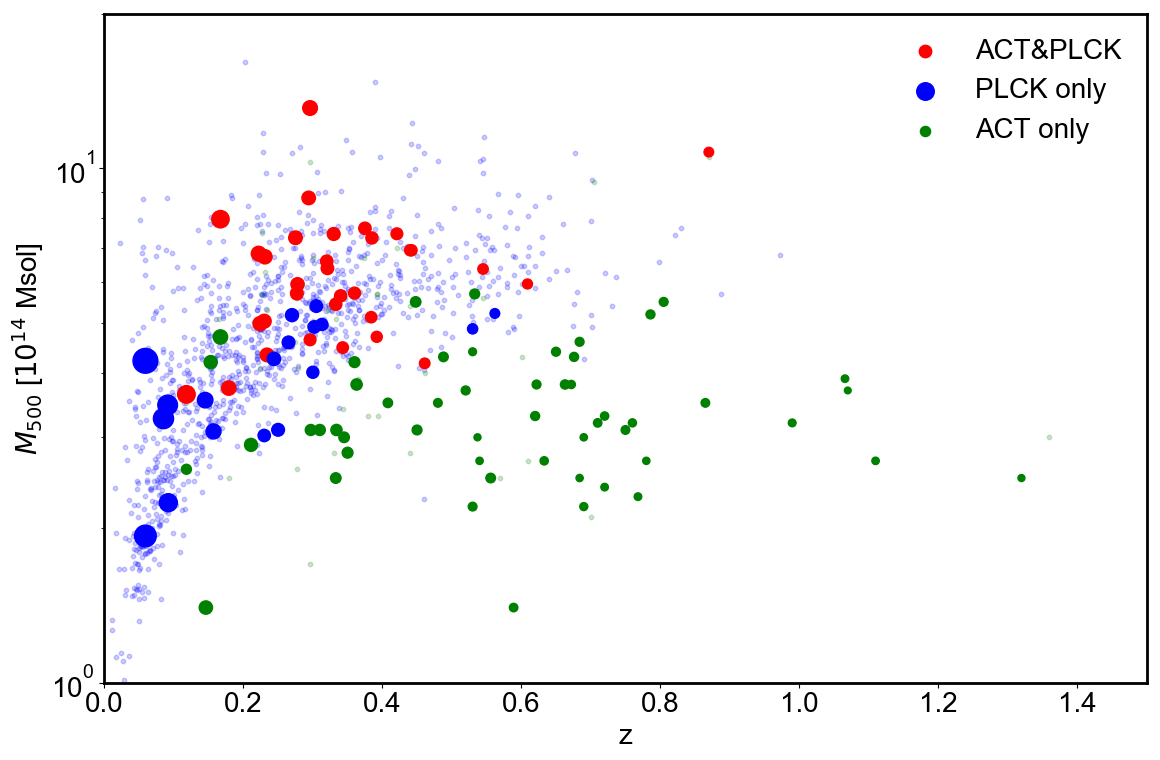}}
\caption{Distribution in mass (in units of $M_{500}$ and redshift of our sample (filled circles) compared with the full {\it Planck} and ACT catalogs (light dots). The colors indicate whether a cluster is detected by ACT (green) or {\it Planck} (blue) or both (red). The size of the filled circles is proportional to the cluster size, $\Theta_{500}$.}
\label{fig:Mz}
\end{center}
\end{figure}

Figure \ref{fig:Mz} shows the distribution of the 119 clusters considered in this study in the mass-redshift space. The cluster masses, $M_{500}$, are those reported in the ACT and {\it Planck} catalogs. Colors indicate whether a cluster is detected by ACT (green filed circles) or {\it Planck} (blue filled circles) or both (red filled circles) and size of the symbol is proportional to the angular size.


\subsection{Significance of tSZ detections} \label{sec:sn}

\begin{figure}[!th]
\begin{center}
\resizebox{\hsize}{!}{\includegraphics{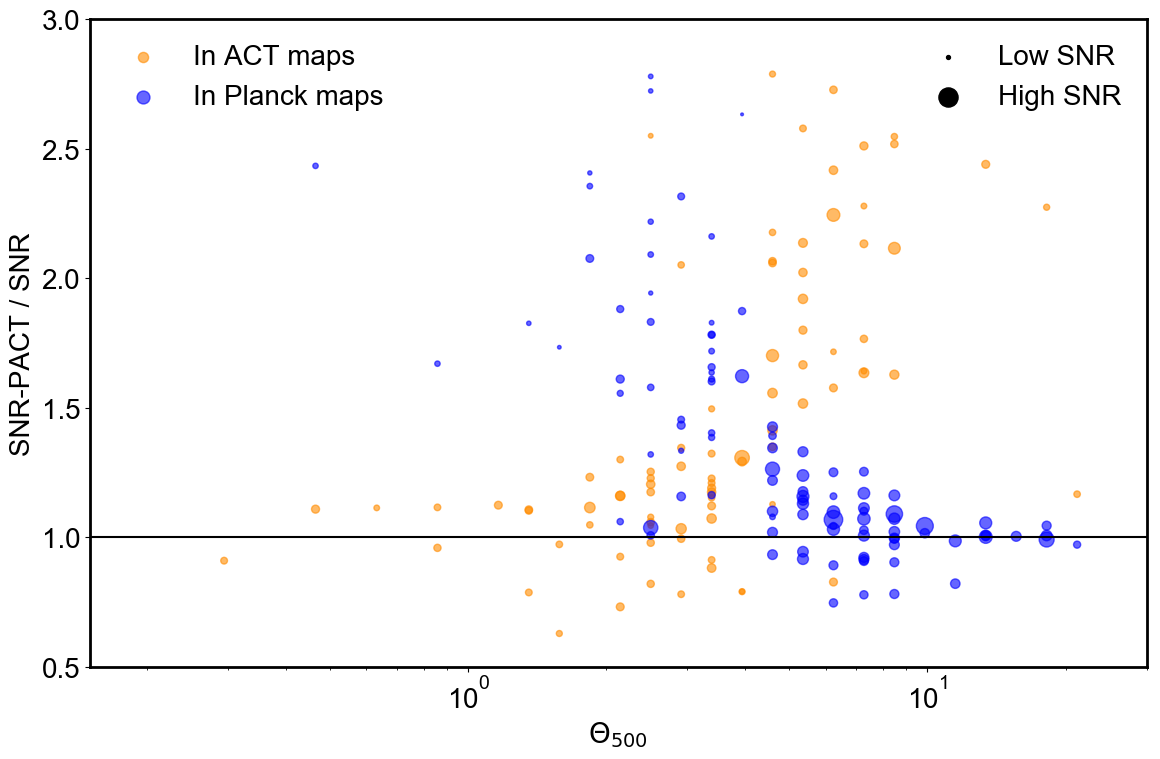}}
\caption{Comparison of the S/N values for the clusters in the ACT, {\it Planck} and PACT maps as a function of angular size. The ratio of the S/N obtained from PACT to the S/N obtained from the native ACT data (orange dots) or {\it Planck} data (blue dots) is plotted versus the angular size of the cluster. Each dot stands for one of the 119 clusters reported in {\it Planck} or ACT catalogs with size of symbol proportional to the PACT S/N. }
\label{PACT_SN_comp}
\end{center}
\end{figure}

For each cluster of the union sample displayed in Fig. \ref{PACT_SN_comp} as a dot, we estimate with the MMF technique the size, the flux, and the S/N at the reported positions. We leave the discussion on the size-flux degeneracy to the next section and focus only on the S/N. The cluster properties are computed with the same MMF from the ACT data (orange dots) and {\it Planck} data (blue dots), i.e. two and six frequency channels respectively, and from the PACT dataset, i.e. the ensemble of eight frequency channels. We show in Fig. \ref{PACT_SN_comp} the ratio of the S/N obtained from PACT to the S/N obtained from the native dataset (ACT or Planck) with the same MMF as a function of the cluster size.  

Figure~\ref{PACT_SN_comp} highlights the improvement on the detection signal-to-noise ratio both for extended clusters and more compact clusters, although a few clusters have a lower S/N when combining ACT and {\it Planck} data. We first notice that for most of the ACT and {\it Planck} clusters the S/N improves when we use the PACT combined data rather than the native ACT({\it Planck}) data to estimate the signal-to-noise ratio. The improvement can reach up to a factor three higher than the ACT({\it Planck}) native S/N. In the case of ACT (orange dots), the highest improvements correspond to clusters with the largest sizes. In these cases, the PACT combination provides the large angular scales of {\it Planck}, missing from the ACT data, which also filter out the CMB and astrophysical contamination. In the case of {\it Planck} (blue dots), the highest improvements in S/N are reached for clusters with small angular sizes. The PACT dataset provides the small scales of ACT that are missing in {\it Planck} low-resolution data. For clusters with intermediate sizes, the average improvement in S/N induced by the use of the PACT dataset is of order 50\%. We also note from Fig. \ref{PACT_SN_comp} a few clusters (less than 10\% of the full sample) exhibit smaller S/N values when the PACT dataset is used. Such a downward fluctuation of the S/N despite the use of more data was already noted by \cite{planck2016-l01} (while the S/N improved for most SZ sources about 23\% had their S/N reduced below the detection limit of $4.5\,\sigma$). These downward-fluctuated S/N were associated with weak detections likely to be Eddington biased above the threshold. We note in the present analysis of PACT dataset that the clusters with reduced S/N are low significance ones. The modification of the background noise in the aggregated PACT dataset and the low significance of the tSZ detections result in downward-fluctuated detections. For the most nearby and thus extended clusters (such as A3395 at redshift $z=0.05$), the background noise used in the MMF technique is overestimated due to the cluster signal at large scales. A dedicated MMF analysis of such extended sources would be needed. 


\subsection{Size and tSZ flux relation}\label{sec:flux}

\begin{figure}[!th]
\begin{center}
\resizebox{\hsize}{!}{\includegraphics{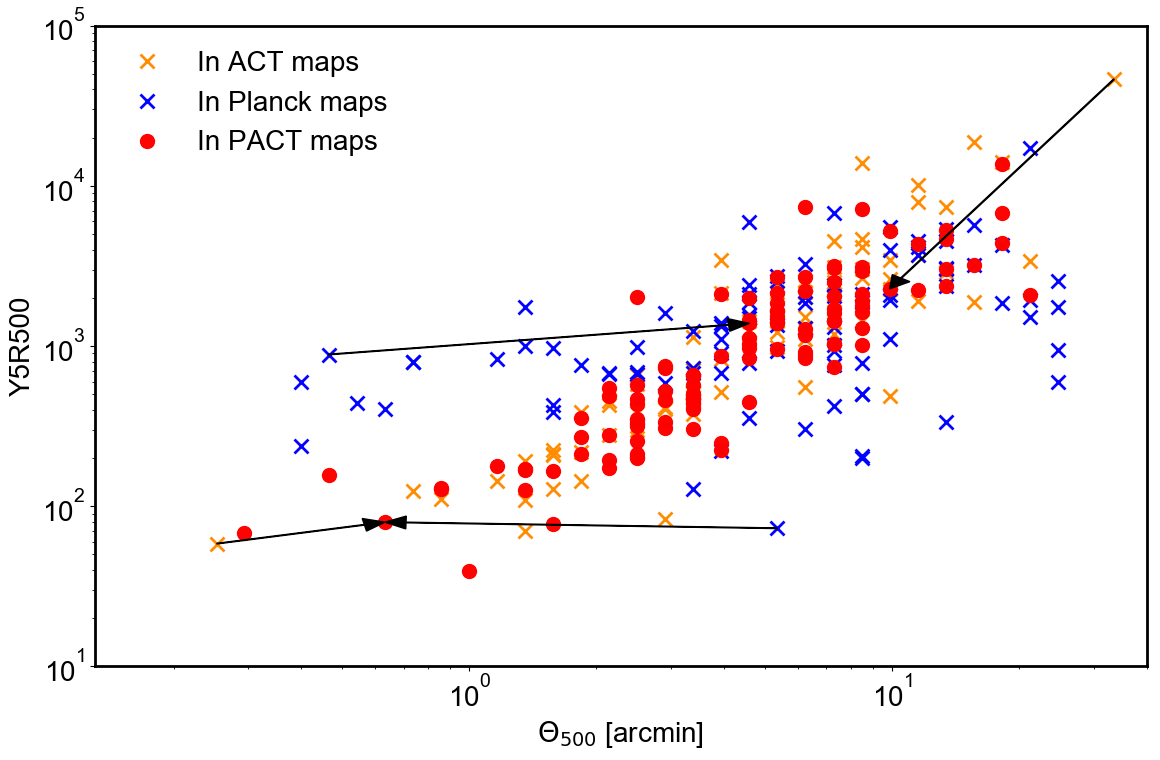}}
\caption{Comparison the $Y$ and $\Theta$ values for the 119 tSZ clusters in the ACT, {\it Planck} and PACT maps. Arrows show how the size and flux change when estimated in native ACT/{\it Planck} maps only or using the full PACT dataset. }
\label{PACT_Y_comp}
\end{center}
\end{figure}

The MMF methods adjusts simultaneously the size, $\theta_{500}$, and the integrated Compton parameter, $Y_{500}$, on the data. We show in Fig. \ref{PACT_Y_comp}, the tSZ flux and the size measured with the same MMF technique at the position of all the 119 clusters detected by ACT and {\it Planck} in the ACT footprint. The red dots show the outputs from the MMF applied to the combined PACT dataset. The orange and blue crosses are for MMF values obtained from the ACT and {\it Planck} data, respectively.

We first note the large dispersion in the $Y_{500}$-$\theta_{500}$ relation for the {\it Planck} case (blue crosses) which is due to the size-flux degeneracy. As was shown \citep{planck2011-5.1a,planck2013-p05a}, both quantities estimated simultaneously with the MMF filter are highly correlated. This degeneracy is particularly important in the case of low resolution data such as {\it Planck} for which clusters are only marginally resolved. In the case of ACT (orange crosses), the resolution is much higher and the cluster sizes are better estimated, on average.
We also note, in Fig. \ref{PACT_Y_comp}, that the tSZ flux expressed in terms of the $Y_{500}$ parameter can be overestimated both in {\it Planck} and in ACT. In the ACT case, $Y_{500}$ is overestimated for extended clusters. This is due to the fact that ACT lacks the large scales and is thus the data are contamination mostly by CMB and dust emission. In the {\it Planck} case, $Y_{500}$ values are overestimated for small-size clusters unresolved by Planck. In addition at the smallest sizes, we notice that $Y_{500}$ values show a plateau which is due the noise in the {\it Planck} maps. We can see the same kind of plateau for the case of ACT but at lower amplitudes (noise). 

\begin{figure*}[!th]
\begin{center}
\includegraphics[scale=0.15]{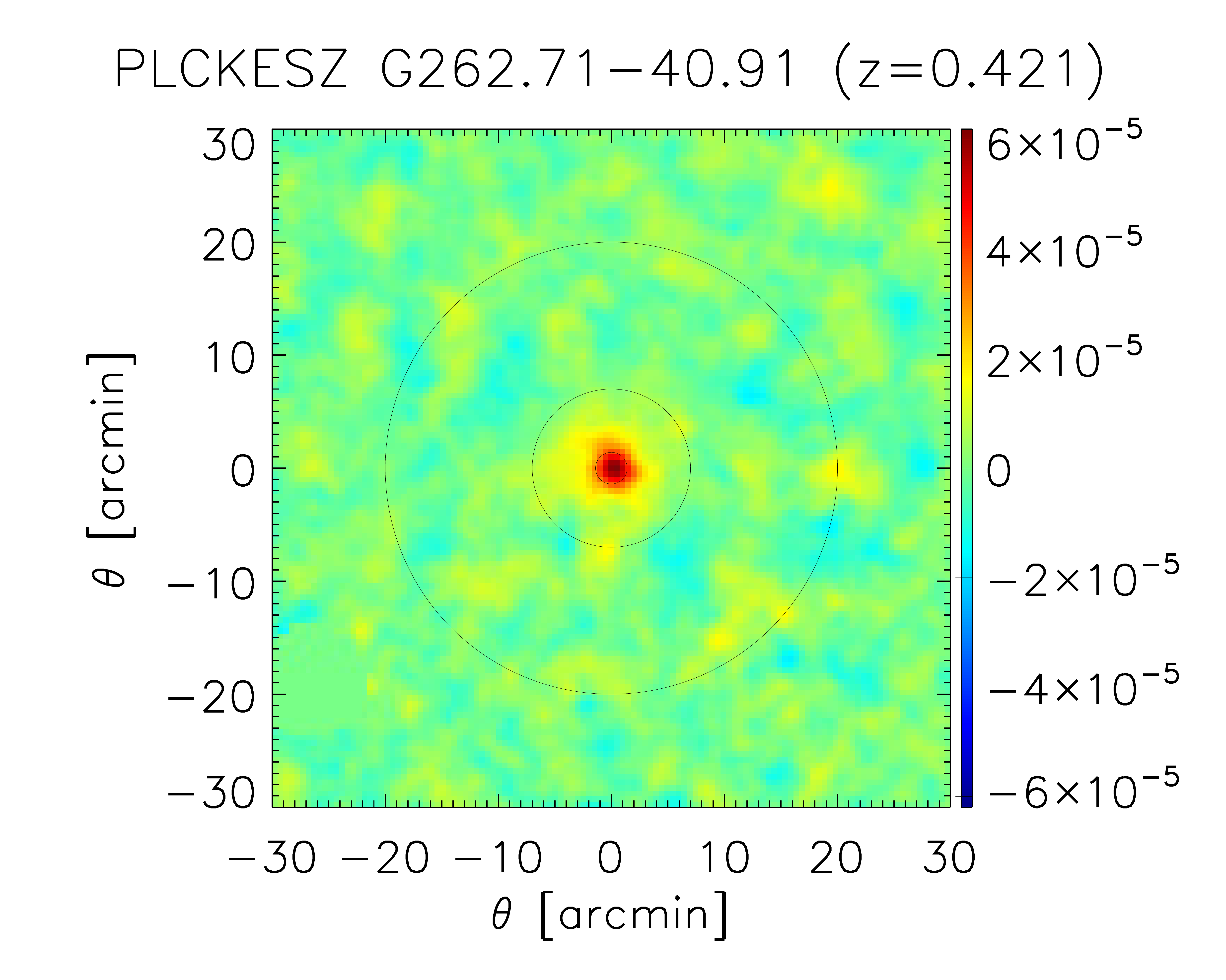}
\includegraphics[scale=0.15]{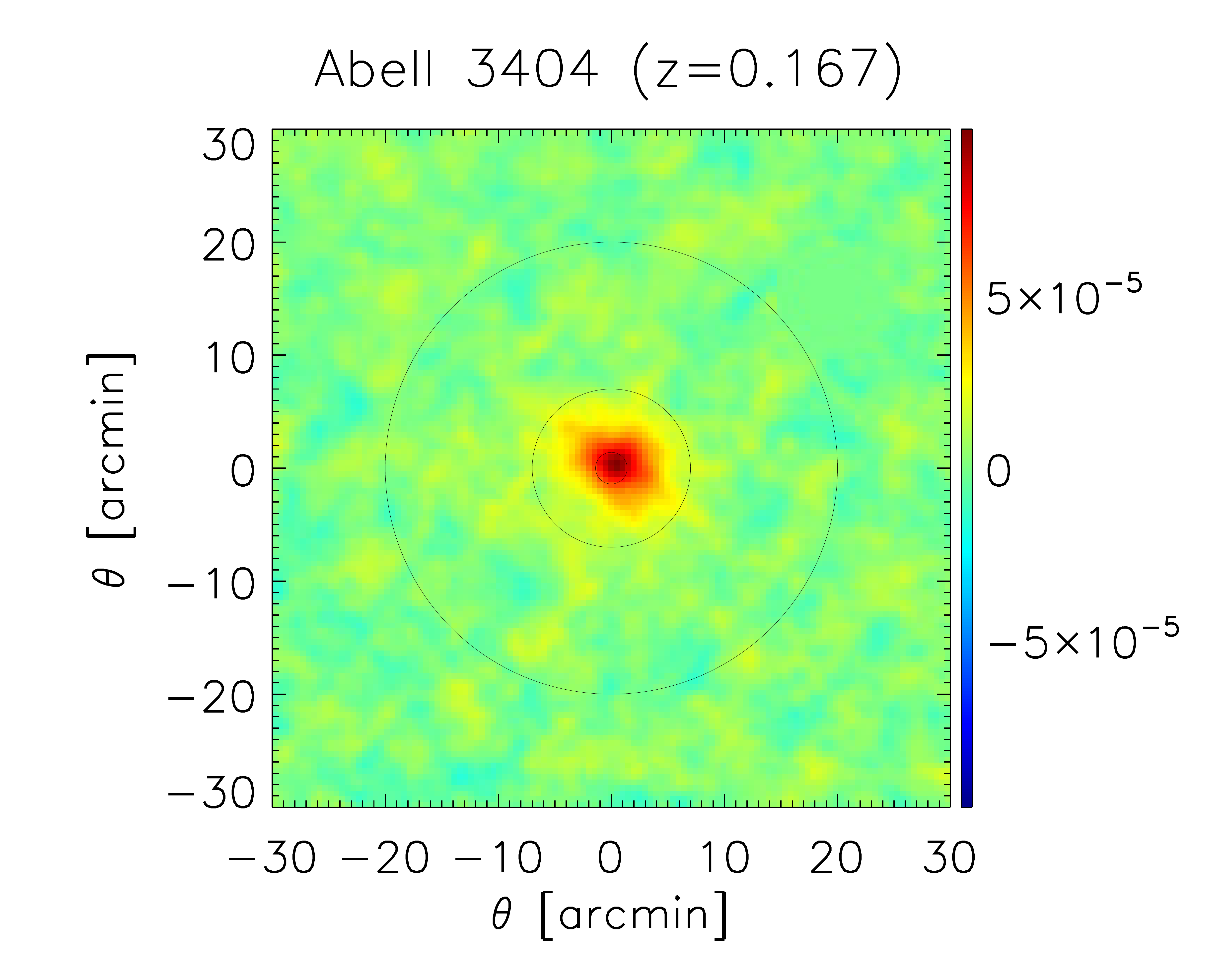}
\includegraphics[scale=0.15]{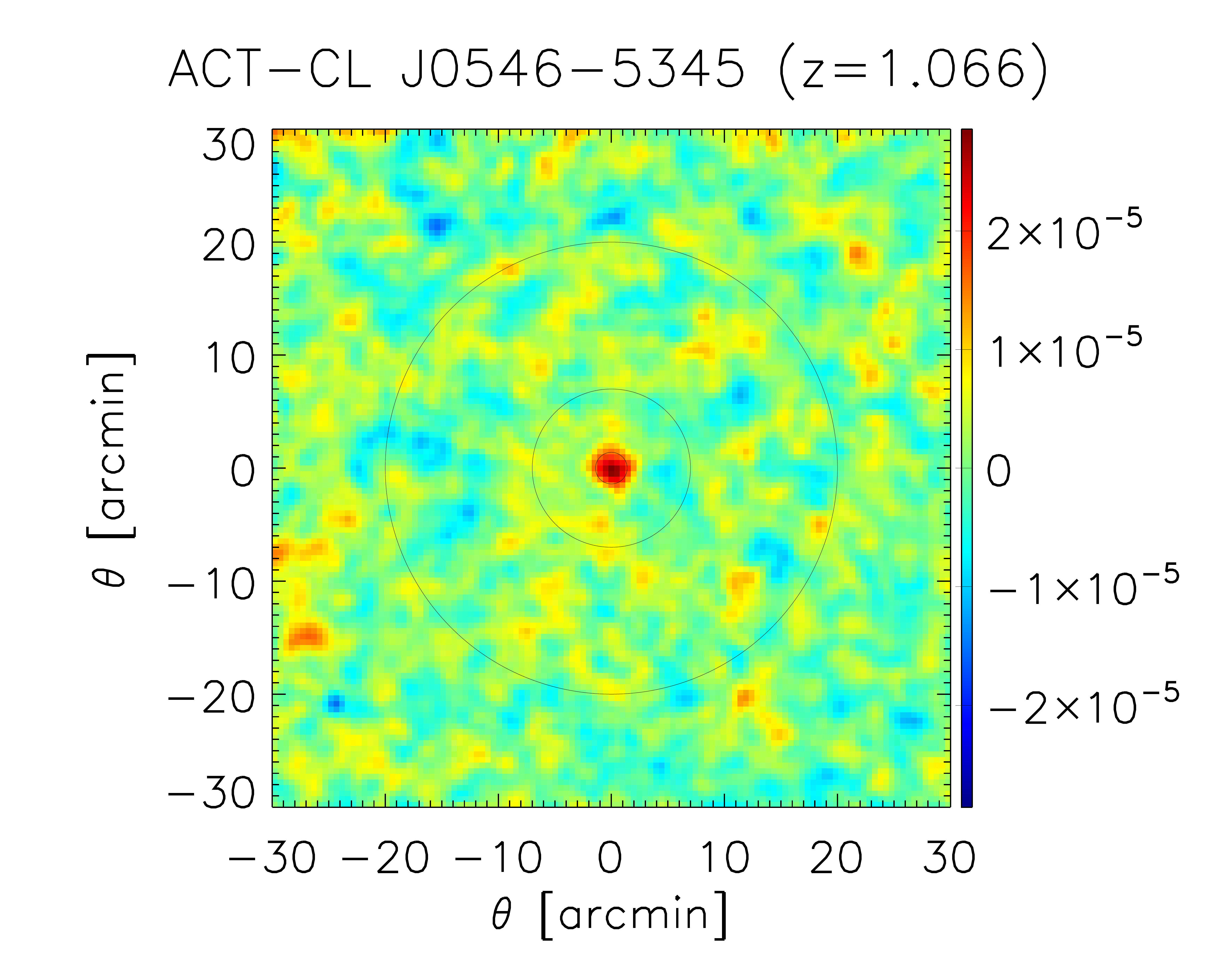}
\includegraphics[scale=0.15]{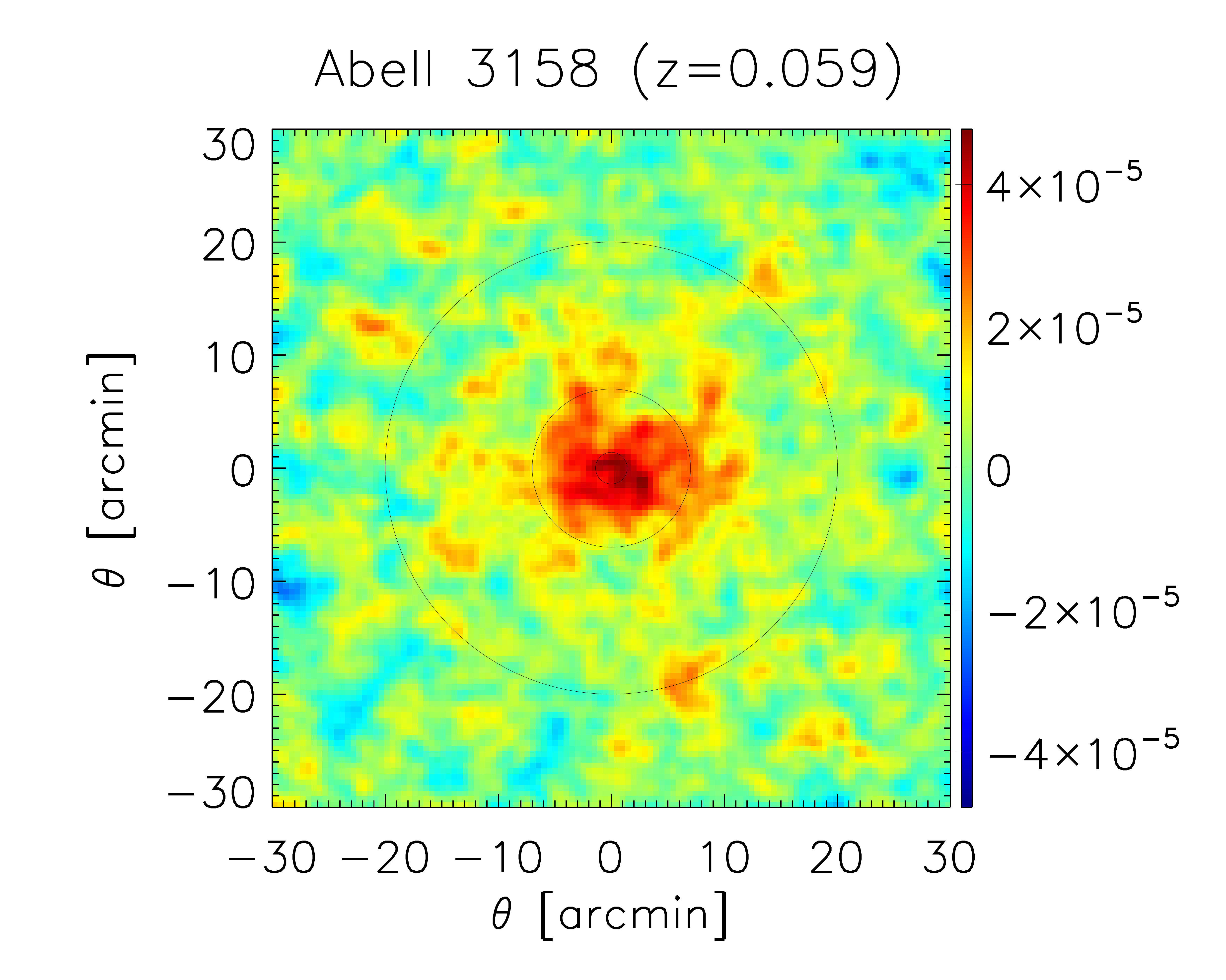}
\caption{For four selected galaxy clusters, cut-outs of the tSZ signal from the reconstructed PACT $y$-map. The inner black circle is for 1.4 arcmin (ACT resolution), the middle black circle represents the {\it Planck} resolution of 7 arcmin, and the outer black circle is drawn for a radius of 20 arcmin.}
\label{fig:pmap}
\end{center}
\end{figure*}

\begin{figure*}[!th]
\begin{center}
\includegraphics[scale=0.15]{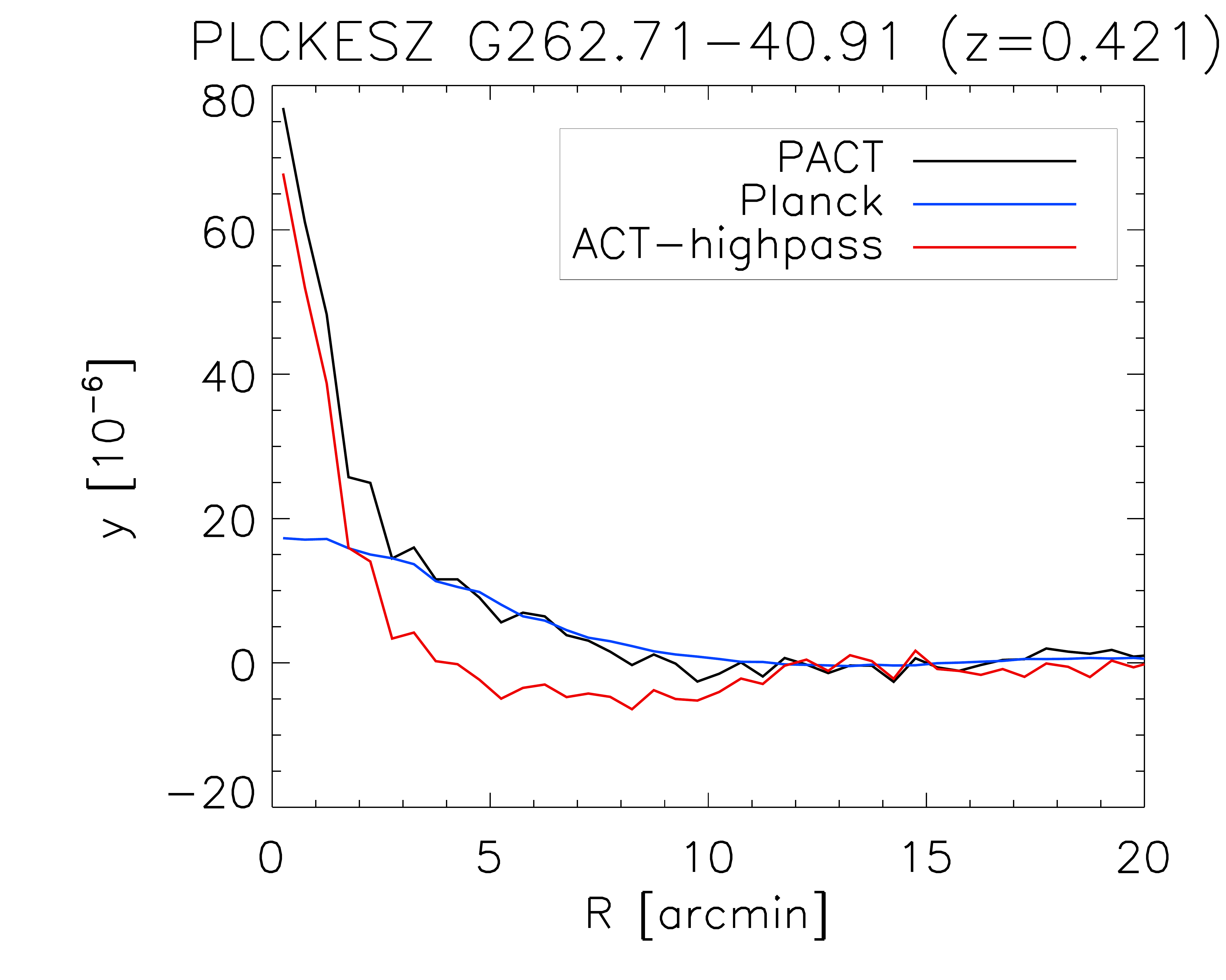}
\includegraphics[scale=0.15]{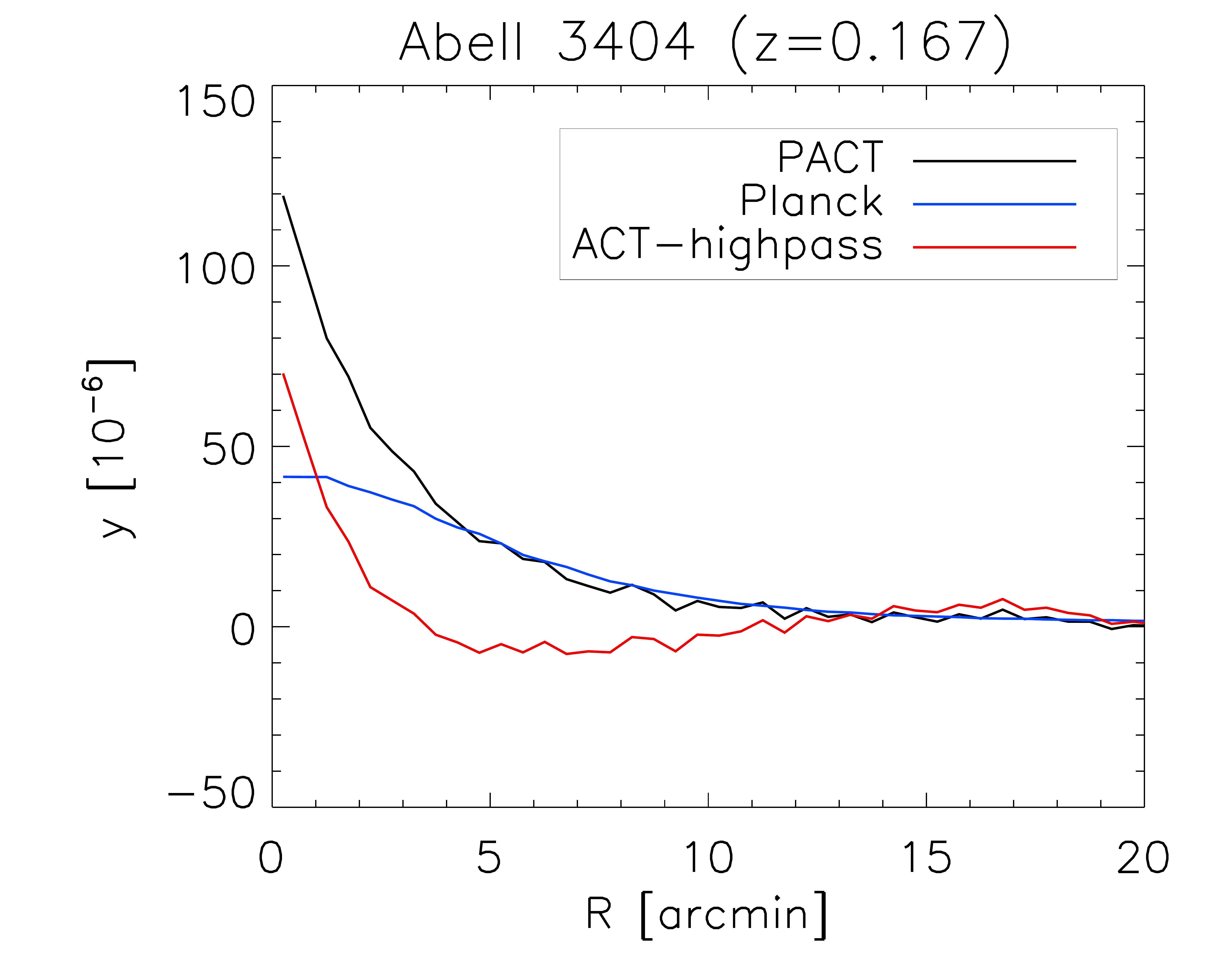}
\includegraphics[scale=0.15]{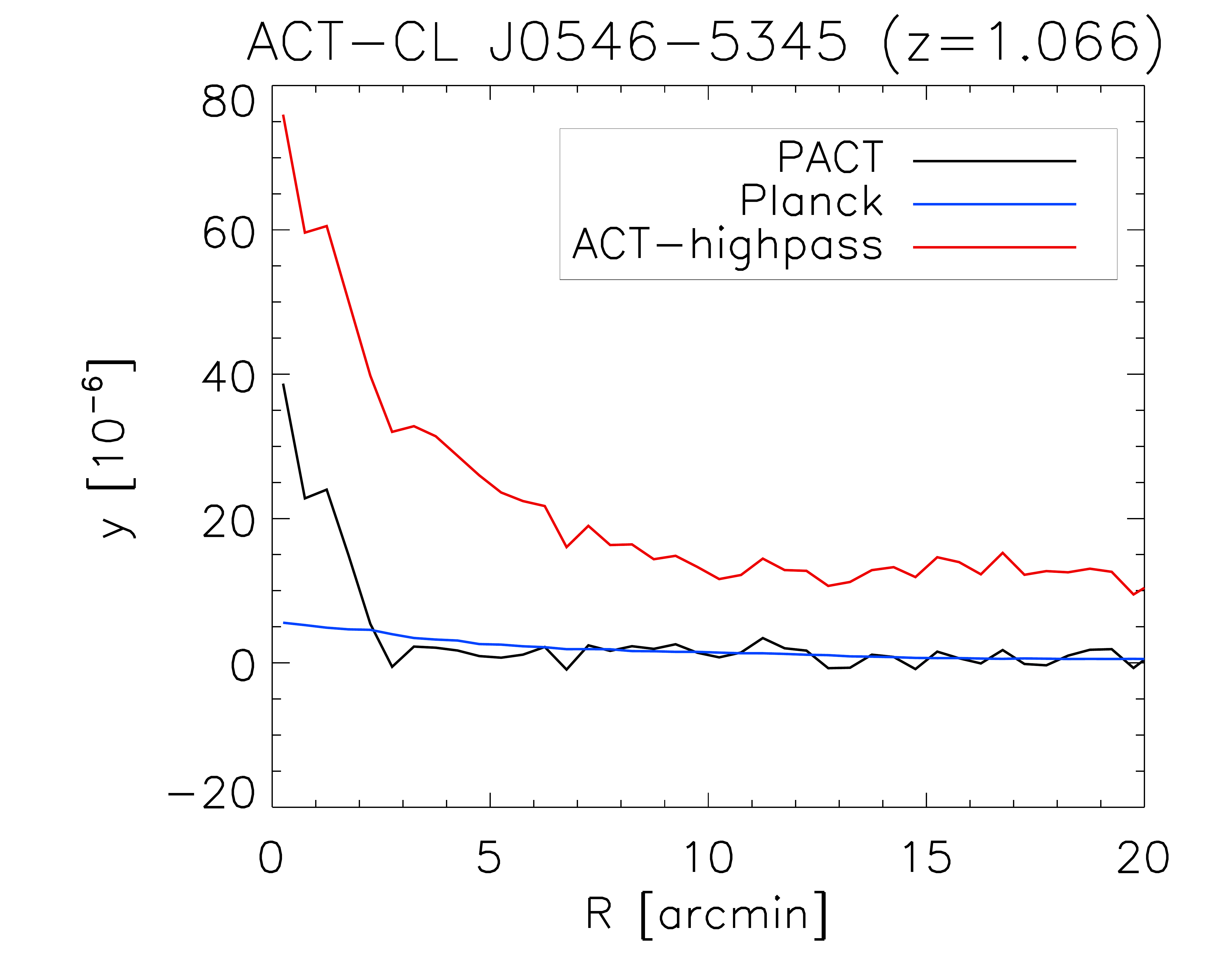}
\includegraphics[scale=0.15]{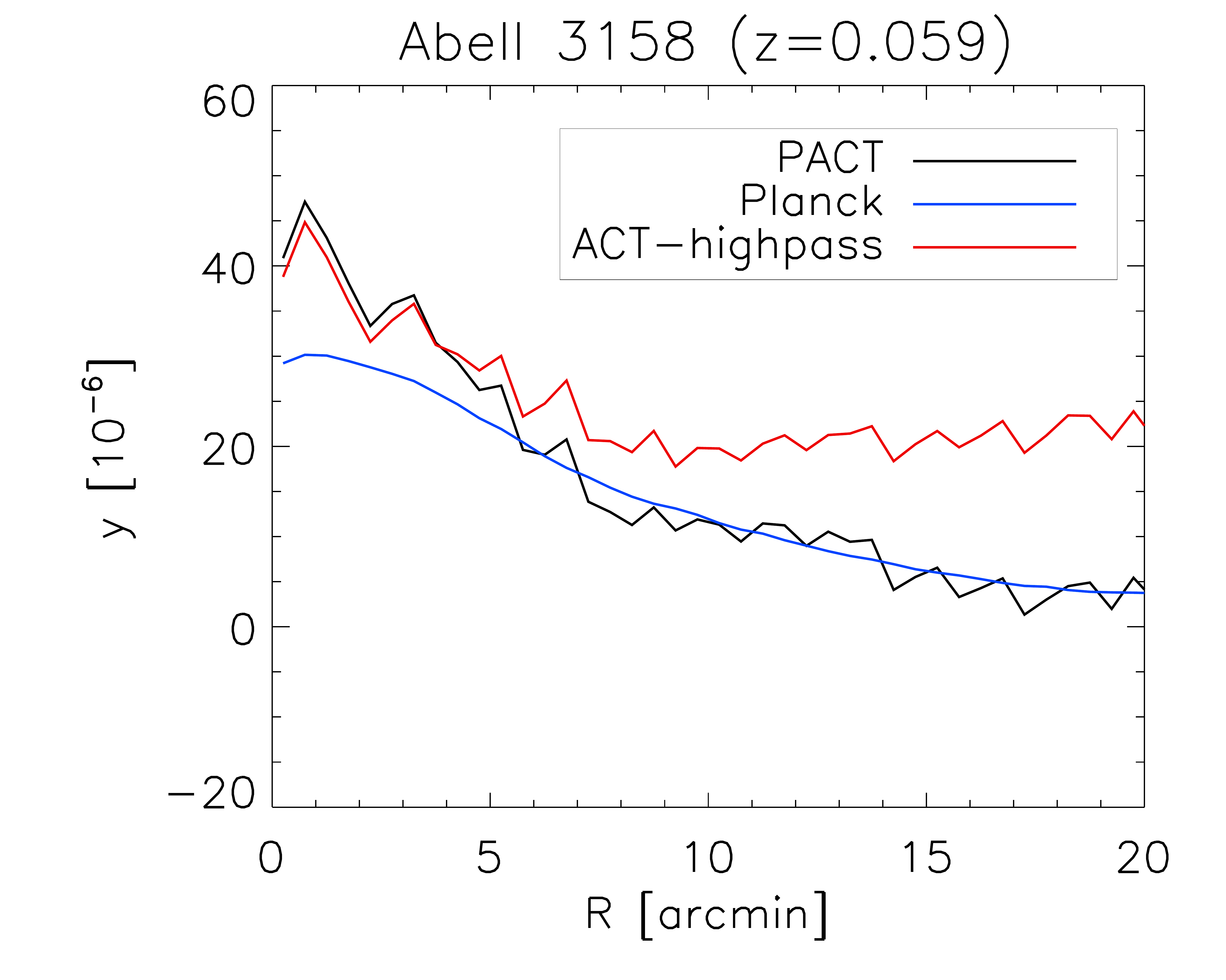}
\caption{For the four clusters in Fig.~\ref{fig:pmap}, comparison of the radial tSZ profiles obtained from the PACT $y$-map (black), from the {\it Planck} $y$-map (blue), the 148GHz-ACT high-pass filtered map (red).}
\label{fig:prof}
\end{center}
\end{figure*}

When we use the combined PACT dataset for our MMF analysis (red dots) this induces an improvement over both ACT and {\it Planck} alone. This is triggered by the improved measurement of the sizes.
Extended clusters in ACT benefit from the {\it Planck} large scale signal and multiple frequency which reduces the contamination and hence the $Y_{500}$ estimate (orange crosses and arrows pointing to the left in Fig. \ref{PACT_Y_comp}). Unresolved clusters in {\it Planck} benefit from the small scales of ACT which permits to actually measure the cluster size and in turn the integrated Compton parameter $Y_{500}$ (blue crosses and arrows pointing to the right). Finally, for the smallest sizes the dominant contribution of the ACT data in the PACT combination reduces the noise level and further improves the tSZ estimates.
The joint analysis of ACT and {\it Planck} with the MMF method hence reduces the size-flux degeneracy leading to a much smaller dispersion of the $Y_{500}$-$\theta_{500}$ relation.


\subsection{tSZ profiles}

We have shown, in Sect. \ref{sec:map}, that the combination of ACT and {\it Planck} channels maps allows us to reconstruct a PACT $y$-map that benefits on the one hand from the ACT high resolution and low noise, and on the other hand from {\it Planck} large scale modes and multiple frequencies. In this way, the reconstructed PACT $y$-map has better resolution and reduced noise and astrophysical contamination. We now illustrate how the combined PACT map improves the measurement of the tSZ profile on individual clusters of galaxies. 

Dedicated analyses by Crichton et al. (in prep) and Santiago-Bautista et al. (in prep) focus on the derived pressure profile from PACT data. Here, we select only four clusters with different angular sizes in order to exhibit the advantage of the combined PACT $y$-map over the use of ACT or {\it Planck} maps independently. Namely, we choose a nearby and very extended cluster: Abell 3158 at $z=0.059$; a compact cluster of size smaller than the {\it Planck} beam: Abell 3404 at $z=0.167$; an intermediately-high redshift cluster: PLCKESZ G262.71-40.9 at $z=0.421$; and finally a high redshift cluster at the resolution limit of Planck: ACT-CL J0546-5345 at redshift $z=1.066$. 

For each cluster, we show in Fig. \ref{fig:pmap} the tSZ $1^\circ\times1^\circ$ patch from the PACT $y$-map, centered at the cluster position. The black circles displayed in the images are drawn for radii of 1.4, 7 and 20 arcmin, from the inner to the outer circle. The associated radial Compton parameter profiles, derived from the PACT $y$-map, are shown in Fig. \ref{fig:prof} as the black lines. We also plot the profiles obtained from the {\it Planck} $y$-map (blue lines) together with those obtained from the high-pass filtered ACT-148~GHz  map (red lines). 
This figure  exhibits the complementarity in
PACT of both ACT and {\it Planck} data,
respectively. This is particularly well illustrated in
Fig. \ref{fig:prof} upper right panel where the smallest-scale tSZ signal is provided by ACT and the outskirt of the tSZ signal by {\it Planck}. 

\section{Conclusion}\label{sec:ccl}

We have built a composite dataset from the ACT and {\it Planck}-HFI channel maps and their associated noise maps. This dataset was used to perform an optimised reconstruction and extraction of the tSZ signal. In practice, we have followed two approaches. The first one is source-oriented and hence focused on the detection of tSZ clusters using a Matched Multi-Filter detection method. The second approach is based on the reconstruction of a tSZ $y$-map using an  ILC technique. 
For either approach, we have shown that the composite PACT dataset benefits from the high resolution of ACT  and the large scale modes and large number of frequency channels of {\it Planck}.

For the tSZ cluster detection, we have shown that we recover all clusters detected in the individual datasets of {\it Planck} and ACT. In addition, we detect new tSZ sources some of which associated with {\it bona-fide} clusters at redshifts up to 0.93, that were neither reported by {\it Planck} nor by ACT independently.  We have also shown that the combination of ACT and {\it Planck} data reduces the size-flux degeneracy and improves the signal-to-noise of detected clusters by up to a factor three with respect to their native ACT or {\it Planck} signal-to-noise. 

For the Compton $y$ parameter map, we have used a scale-dependent optimization of the reconstructed tSZ signal with respect to noise and foreground residuals. We have shown that the resulting PACT y-map reproduces the tSZ signal from arcminute angular scales present in ACT data to the largest  the angular scales captured by {\it Planck}. 
The improvement in the reconstructed tSZ PACT map was also illustrated by the derived Compton $y$ profiles for representative clusters, showing the complementary of ACT and {\it Planck} to reproduce both the inner structure and outskirts of the pressure profile. This new PACT $y$-map will be used for further analyses of the pressure profiles and hot gas distribution in individual structures. 

This first optimal combination of multi-frequency and mutli-experiment data illustrates the advantages of data aggregation for tSZ studies. With the forthcoming increase of publicly available data (e.g., ACTpol, AdvACT, SPTpol, SPT3G), joint studies will allow a synoptic analysis of tSZ over the majority of the extragalactic sky.

\begin{acknowledgements}
This research was performed in the context of the ISSI
international team project "SZ clusters in the Planck era" (P.I. N. Aghanim \& M. Douspis). The authors acknowledge partial funding from the DIM-ACAV and the Agence Nationale de la Recherche under grant ANR-11-BS56-015. The research leading to these results has received funding from the European Research Council under the H2020 Programme ERC grant agreement no 695561. JMD acknowledges support from project AYA2015-64508-P (MINECO/FEDER, UE) funded by the Ministerio de Economia y Competitividad. DC acknowledges support from the South African Radio Astronomy Observatory, which is a facility of the National Research Foundation, an agency of the Department of Science and Technology.
The development of {\it Planck} has been supported by: ESA; CNES and CNRS/INSU-IN2P3-INP (France); ASI, CNR, and INAF
(Italy); NASA and DoE (USA); STFC and UKSA (UK); CSIC, MICINN, JA and RES (Spain); Tekes, AoF and CSC (Finland); DLR and MPG (Germany); CSA (Canada); DTU Space (Denmark); SER/SSO (Switzerland); RCN (Norway); SFI (Ireland); FCT/MCTES (Portugal); and PRACE (EU). This research
has made use of the following databases: the NED and IRSA databases, operated by the Jet Propulsion Laboratory, California Institute of Technology, under contract with the NASA; SIMBAD, operated at CDS, Strasbourg, France; SZ cluster database (szcluster-db.ias.u-psud.fr) operated by Integrated Data and Operation Center (IDOC) operated by
IAS under contract with CNES and CNRS. This research made use of Astropy, the community-developed core Phyton package. 
\end{acknowledgements}


\bibliographystyle{aa}
\bibliography{Planck_bib,pact}

\end{document}